\documentclass[11pt]{article} 

\usepackage[utf8]{inputenc} 

\usepackage{geometry} 
\usepackage{graphicx} 

\usepackage{booktabs} 
\usepackage{array} 
\usepackage{paralist} 
\usepackage{verbatim} 
\usepackage{subfig} 
\usepackage{hyperref}
\usepackage{amsmath}
\usepackage{physics}
\usepackage{braket}
\usepackage{amssymb}
\usepackage{euscript}
\usepackage{mathrsfs}
\usepackage{enumitem}       
\usepackage{dsfont}
\usepackage{authblk} 
\usepackage{url}
\usepackage[round]{natbib}

\usepackage{fancyhdr} 
\usepackage{sectsty}
\allsectionsfont{\sffamily\mdseries\upshape} 

\usepackage[nottoc,notlof,notlot]{tocbibind} 
\usepackage[titles,subfigure]{tocloft} 

\DeclareUnicodeCharacter{211E}{$\infty$}

\usepackage[english]{babel} 

\usepackage[dvipsnames]{xcolor}

\usepackage{amsthm}

\theoremstyle{definition}
\newtheorem{definition}{Definition}

\newcommand{\vN}[1][]{\mathcal{A}^{\text{vN}}_{\text{#1}}}
\newcommand{\Ha}[1]{H_\text{#1}}

\newcommand{\A}[1]{\mathcal{A}_\text{#1}}
\newcommand{\metric}{g_{\mu\nu}}
\newcommand{\modelthree}[3]{\langle #1, #2, #3 \rangle}

\newcommand{\Order}[1]{\mathcal{O}(#1)}

\title{On Algebraic Background-Independence\footnote{Published in \textit{European Journal for Philosophy of Science} 16, 96 (2026), \url{https://doi.org/10.1007/s13194-026-00799-w}}}
\author[1,2]{Luka Vinck}
\author[1,3]{Sebastian De Haro}
 \date{October, 2026} 
\affil[1]{Institute of Physics,
University of Amsterdam, The Netherlands} 
\affil[2]{Dutch Institute for Emergent Phenomena, The Netherlands}
\affil[3]{Institute for Logic, Language and Computation,
University of Amsterdam, The Netherlands}

\begin{document}

\maketitle
\begin{abstract}
It has recently been claimed that algebraic methods can be used to give a background-independent description of semiclassical gravity.  This paper aims to provide an entry point into recent algebraic approaches to quantum gravity and to analyse and critically assess the conception of background-independence here involved. We discuss an algebraic framework for formulating an algebra of observables along the worldline of an observer in a cosmological setting. By ‘gravitationally dressing’ the observables to the observer’s worldline, the reference to the spacetime background can be set aside at the level of the abstract *-algebra. Thus this *-algebra is taken to be independent of any particular spacetime realisation, with different states on the algebra corresponding to different backgrounds. We compare this algebraic framework to three influential notions of background-independence in the literature and argue that none of these fully captures the sense in which the *-algebra is taken to be background-independent. Building on this, we extract a novel notion of algebraic background-independence tied to the representation-independence of the *-algebra. This algebraic conception motivates two further concepts of background-independence tailored to semiclassical theories: semiclassical background-independence and perturbative background-independence. We suggest that these provide useful tools for evaluating the scope and limitations of semiclassical approaches to quantum gravity.

\end{abstract}
\vspace{1cm}

\newpage
\tableofcontents

\newpage
\section{Introduction}\label{introduction}
\label{sec:introduction}
This paper has two main aims: first, to analyse and assess a notion of background-independence that has emerged in recent work in quantum gravity. Second, to provide an entry point into the recent applications of algebraic quantum field theory (AQFT) in quantum gravity that have motivated the discussion of background-independence.

Although it is generally believed that a theory of quantum gravity should be `background-independent', there is, among both physicists and philosophers, no consensus about the precise meaning of background-independence. Broadly speaking, background-independence is usually taken to mean that spacetime is not some absolute structure on which the theory is defined, but is itself a dynamical part of the theory. First explicit usage of the term seems to have appeared in the 1980s, when string theorists started to argue that background fields like the target space metric should be determined dynamically \citep[see e.g.][]{horowitz_purely_1986}. The term was further popularised in the 1990s --- both by \citet{witten_background-independent_1992,witten_quantum_1994,witten_reflections_1996}, who became an influential advocate for the case of background-independence in string theory, and by \citet[e.g.][]{rovelli_discreteness_1995,rovelli_physical_1994}, who took background-independence as a foundational principle in the loop quantum gravity project and criticised string theory for not taking this central lesson from general relativity seriously enough (see also \citealt{smolin_three_2001,smolin_case_2006}).\footnote{For a reply to some of Smolin's criticisms, see \cite[p.~114--116]{de_haro_dualities_2017}.} The resulting polemics are perhaps what put background-independence on the radar of philosophers of physics, who started to analyse and formally define the concept from the 2010s. An overview and a discussion of these definitions are given in a recent book by \citet{read_background_2023}. 

However, even though the term itself may be relatively new, similar ideas have been present from the earliest attempts to quantise gravity. For example, \citet{bergmann_non-linear_1949} sought to extract from general relativity its `essential' lesson with the view of quantising it (Bergmann called this the `skeletonized' theory), and studied the general structure of field theories in which all fields are determined dynamically on the basis of the variation of an action. This reflects the same idea as a recent definition of background-independence in terms of a variational principle \citep{pooley_background_2017}. One of Bergmann's collaborators in Syracuse, James \citet{anderson_principles_1967}, would some time later formulate what is retrospectively seen as an early conceptual definition of background-independence under a different name, namely in terms of `absolute objects'. Around the same time, \citet{misner_gravitation_1973}, in their influential book on gravitation, introduced the slogan `no prior geometry' to capture the distinguishing feature of GR. In other words, what characterises these early ideas on background-independence is that they sought to single out the essential aspect in what GR teaches us about the world and that should be carried over to a theory of quantum gravity.

Recent discussions of background-independence draw on a convergence of three approaches: namely, holography, studies of de Sitter space, and AQFT. For example, recent approaches to AdS-CFT have emphasized that gravitational physics is encoded in a boundary algebra \citep{leutheusser_causal_2023,leutheusser_emergent_2023,leutheusser_subregion-subalgebra_2025}, and newly developed techniques such as entropy bounds, spacetime reconstruction, and quantum error correction, have used the tools of von Neumann algebras and modular theory \citep[for a review, see][]{liu_lectures_2025}. The algebraic approach is particularly well-suited to the definition of causal structures and entropies: for this reason, it seems natural to study quantum gravity in de Sitter-type spacetimes, where each observer has both an associated causal horizon and an entropy. Furthermore, the algebraic approach is predicated on a clear distinction between states and observables, such that the algebra is independent of the state. This is conceptually and technically useful in quantum gravity, because it allows one to focus on defining observables without immediately committing to a specific state. 

However, as \citet{witten_background_2025} argues, there are three challenges in trying to apply the algebraic approach to quantum gravity:
\begin{enumerate}[label=(\roman*)]
\item {\bf Fluctuating spacetime:} The most obvious problem is that an AQFT, as usually formulated, assigns to each open region $U\subset M$ an algebra ${\cal A}_U$, i.e.~$U\mapsto {\cal A}_U$\footnote{This assignment is called a `net of algebras,' see \citet{haag_local_1996}.}. Thus if the spacetime fluctuates, it is unclear what these regions are, and how to define such an assignment.
\item {\bf Background-independence:} In standard AQFT, the algebras depend on the spacetime but are independent of the quantum state: changing the state leaves the net $U\mapsto{\cal A}_U$ unchanged. By contrast, in the presence of gravity, the quantum state of the fields affects the spacetime geometry. Thus, if we require that the algebra should be independent of the state in the AQFT sense, it cannot be defined by reference to a particular spacetime.
\item {\bf Internal observers:} In asymptotically AdS or asymptotically flat spacetimes one assumes that gravity is effectively ``turned off" at the boundary, so that the asymptotic structure can be treated as a non-dynamical boundary condition. This structure can then be used to define a state-independent algebra, which is naturally interpreted as describing the possible measurements of an external observer. In that setting, the observer can be idealized as lying ``at infinity'', outside the gravitating system, and the associated algebra can be defined with respect to the asymptotic region, independent of the interior state. By contrast, in a closed universe, such as de Sitter-type spacetimes (i.e.~with a positive cosmological constant), there is no `external point of view' to support such an observer, and it is unclear how the algebra should be interpreted in terms of observations.
\end{enumerate}

Challenge (i) is part of the general problem of how to quantize gravity: an interesting issue that, in the semiclassical settings that we will discuss in this paper, we can mostly bracket. The upshot of (ii) and (iii) for the application of AQFT to quantum gravity is that, in de Sitter-type spacetimes, just as we cannot attach observables to asymptotic regions, we also cannot assign algebras to regions in the standard way. 

This shows the interconnectedness between the problem of defining observers and the problem of background-independence in quantum gravity.  Witten's argument in (ii) suggests an interesting reinterpretation of background-independence as the {\it independence of the algebra from the state}. For if, in line with AQFT practice, we require that the algebra be state-independent (as we do, e.g.~when we are interested in representation-independent statements), then in a gravitational context it must also be independent of any particular classical spacetime geometry, since an algebra defined by reference to a given spacetime would itself vary with the state. Thus (ii) does not require background-independence {\it by fiat}: rather, it extracts a notion of background-independence from the algebraic requirement of state-independence of the algebra. This is an important reason why we think that this approach is likely to cast light on the nature of background-independence. 

At the same time,  in a closed universe, any observer must be modelled internally, as part of the gravitating system, and its algebra seems to depend not only on the state of the system, but also on the observer's own state. Background-independence in the sense of (ii) then amounts to the requirement that the observer's algebra is independent of both the state of the system and its own state.  This highlights the relation between the notion of an observer and background-independence in quantum gravity. 

In a recent paper, \citet{chandrasekaran_algebra_2023}---commonly referred to by their initials CLPW---followed by two papers by \citet{witten_algebras_2024,witten_background_2025}, proposed a novel approach to this question in de Sitter-type spacetimes in the semiclassical limit. They constructed an algebra of observables along the worldline of an observer by `gravitationally dressing' (in a sense to be specified in Section \ref{sec:gravdressing}) the observables to the worldline. The algebra can then be interpreted as containing the observables available to an observer on that worldline. 

The advantage of this approach is that it addresses the challenges posed in (i)-(iii). Namely, by defining the algebra with respect to an observer's worldline $\gamma$, we no longer need to refer to a specific region of spacetime. The algebra is generated by fields localised along $\gamma$, and interpreted as describing the observations that this observer can make. So (i) is addressed by formulating an assignment $\gamma \mapsto \mathcal{A}_\gamma$. Of course, $\gamma$ will depend on the specifics of the manifold $M$ to which it maps, but the point is that the definition itself is invariant under changes of the spacetime, and is expected to make sense also when spacetime starts to fluctuate \citep{witten_algebras_2024}. 

Making the construction of the algebra thus independent of the specific structure of spacetime opens the door for a more background-independent formulation: it suggests a more general framework that can in principle be used for any choice of background and is background-independent in the sense of (ii). \citet{witten_background_2025} argues that to get background-independence we should see the algebra as an abstract operator product algebra\footnote{An operator product algebra is not actually an algebra but a set of operators with an operator product expansion \citep[see][]{holland_operator_2013}.} and not as an algebra of Hilbert space operators, since any Hilbert space representation will depend on a particular choice of $M$ and $\gamma$. In that case, a state on the algebra can be associated to a particular choice of $M$ and $\gamma$, and the state-independence of the underlying algebra implies a form of background-independence. 

Also, by explicitly incorporating an observer, CLPW's construction directly addresses the challenge in (iii). The algebra can be interpreted unambiguously as the algebra of observables available to that observer. CLPW's solution to the interconnected problems of observers and background-independence in quantum gravity is thus to directly ``anchor'' the algebra to an observer's worldline as opposed to some fixed spacetime structure.

The aim of the present paper is to review the background-independence of the algebraic formalism proposed by CLPW and to qualify Witten's claims of background-independence.\footnote{Recently, \citet{liu_towards_2025} has also proposed, and illustrated with examples, that the *-algebra structure is background-independent, while various states that can be defined using the GNS construction give rise to different geometries and different asymptotics. Other recent work that builds on Witten's claims of background-independence include \citet{krishnan_state-independent_2024, bak_janus_2024, blommaert_absolute_2025,klinger_theory_2026}.} We will argue that the algebraic framework of \citet{witten_background_2025} should not be considered fully background-independent. For, even though there is an underlying algebraic structure (the operator product algebra or *-algebra) that is background-independent in the sense of (ii), 
to calculate most physical quantities of interest we still have to pick a particular Hilbert space representation for the algebra, associated to a particular state and spacetime background, so that the algebra closes to a von Neumann algebra. For a given choice of background, states in the associated Hilbert space describe $\Order{1}$ perturbations of the input state. These are the dynamics of quantum fields on a fixed spacetime background. Apart from any possible back-reaction effect, the background itself is therefore not dynamical and the theory cannot be considered background-independent in the more conventional sense of the word. 

Despite this, we argue that the framework can be considered background-independent, in a sense that is not captured by existing definitions: it is not so much about the background being dynamical, but about there being some underlying algebraic structure that does not rely on a spacetime formulation at all, and from which different spacetimes can be reconstructed. To explicate and assess this idea, we will formulate a new notion of background-independence, based on (ii), that serves to fill this gap. While this notion is motivated by and modelled on the algebraic framework under consideration, we argue that it applies more generally to semiclassical or linearized approaches to quantum gravity. 

The organization of this paper is as follows. In Section \ref{sectionCLPW}, we first briefly motivate and review the application of AQFT in quantum gravity before introducing CLPW's framework for gravitationally dressing an algebra of observables to the worldline of an observer. Section \ref{section_definitions} contains a short review of three main notions of background-independence in the philosophical literature. In Section \ref{section_pbi}, which includes the main results of this paper, we assess the background-independence of CLPW's framework and introduce three novel notions of background-independence based on the algebraic approach. Finally, Section \ref{sec:conclusion} contains the conclusion. 

\section{An Algebra of Observables Along the Worldline of an Observer}
\label{sectionCLPW}
\subsection{Why Algebraic QFT?}
\label{section:whyaqft}
The adaption of QFT to curved spacetime settings in a semiclassical gravitational setting has led to remarkable insights in quantum gravitational effects. Perhaps most well known of these is Hawking's discovery  that black holes emit quantum radiation and evaporate \citep{hawking_particle_1975}. However, even though QFT can in principle be formulated in arbitrary backgrounds, its formulation for a general spacetime is not straightforward \citep{witten_why_2022}. 

The canonical approach to QFT depends crucially on the Hilbert space: we usually assume that there is a Hilbert space for the states of the system that is unique up to unitary equivalence.\footnote{Two Hilbert space representations are said to be unitarily equivalent if there exists a unitary map from the operators on one Hilbert space to the operators on the other. Unitary equivalence is usually interpreted as a sufficient condition for physical equivalence. For a discussion, see e.g.~\citet{clifton_are_2001}.} However, the construction of the Hilbert space requires us to choose a time direction \citep[see e.g.][]{wald_quantum_1994}. Different Hilbert space constructions can be interpreted as corresponding to different observers moving through spacetime, each of which defines `time' according to their own proper time. Such observers will in general have different notions of vacua and particles. A well-known example is Minkowksi vs.~Rindler foliations of flat space, and their corresponding Hilbert spaces \citep{unruh_notes_1976,fulling_nonuniqueness_1973,davies_scalar_1975,earman_unruh_2011}. 

In Minkowski space, this is not really a problem: it turns out that the Hilbert spaces constructed by different inertial observers are all unitarily equivalent --- they see the same vacuum and the same particles \citep[see][]{ruetsche_interpreting_2002}. Hence the symmetries of Minkowksi space allow us to select a preferred class of unitarily equivalent Hilbert spaces. It is for this reason that, in axiomatic approaches to QFT such as that of \citet{wightman_hilberts_1976}, the existence of a Poincaré invariant vacuum is elevated to an axiom. 

However, in a general spacetime there are no symmetries to appeal to, and so there is no way to use symmetry considerations to single out a preferred Hilbert space representation that would be shared by all (locally) inertial observers. The problem then is that in principle there exists an infinite number of unitarily inequivalent Hilbert space representations of the algebra. This is a consequence of the fact that the Stone-von Neumann theorem, which in quantum mechanics guarantees that all representations of the canonical commutation relations (CCR) are unitarily equivalent to the Schrödinger represention, fails to generalize to the infinite number of degrees of freedom that characterizes QFT \citep[for a discussion, see][]{summers_stone-von_2001}. And without symmetries, so long as we do not want to commit to the perspective of some arbitrary observer, we have no a priori ground to prefer one equivalence class of Hilbert space representations over the other.\footnote{There are however physical arguments to put a condition on the short-distance singularity structure of the states, known as the Hadamard condition, restricting their ultraviolet behaviour. In closed universes, where there is only a finite number of infrared modes, the restriction of the ultraviolet modes by the Hadamard condition means that there is effectively only a finite number of degrees of freedom in the theory, and therefore that the Stone-von Neumann theorem holds. So in that case, imposing the Hadamard condition gives us a preferred unitary equivalence class of Hilbert space representations. In  universes with non-compact Cauchy surfaces, imposing the Hadamard condition means that states belonging to different unitary equivalence classes of Hilbert space representations can be distinguished only by their far infrared modes, but are locally indistinguishable: see \citet{wald_quantum_1994}.} In that case, it does not make sense to speak of ``the'' Hilbert space of the theory. This requires us to take a slightly different approach to QFT.

A different perspective offered by AQFT follows from the observation that, even though we may have many inequivalent Hilbert space representations of the theory, the algebra generated by the operators on these Hilbert spaces is always the same. It therefore proposes to see this algebra as the fundamental structure of the theory rather than the Hilbert spaces on which these operators act. Concretely, this means that after promoting the classical (smeared\footnote{The $\phi(x)$ are operator valued distributions and can be integrated against a test function over some local spacetime region $U$ to get a well-defined operator: $\phi(f)=\int_Ud\mu\phi(x)f(x)$. To get sensible operators, it is necessary to smear the fields along a timelike direction \nocite{witten_algebras_2024}\citep[see also Witten, 2024]{borchers_field_1964}. \label{note:smeared}}) fields $\phi^{(i)}(f)$ to operators and imposing the CCR, we interpret these operators as generating an abstract *-algebra $\mathcal{A}$ and not as operators acting on a specific Hilbert space, as we would normally do in the canonical quantization procedure. We thus work with the abstract algebra and refrain from representing it on a Hilbert space. Instead of taking states to be vectors on a Hilbert space, states are defined as {\it functionals} on the algebra, i.e.~as maps $\psi:\mathcal{A}\to \mathbb{C}$, that are \textit{positive} in the sense that $\psi(A^*A) \geq 0$ for all $A\in\mathcal{A}$ and \textit{normed} such that $\psi(\mathbf{1}) =1$. These are interpreted as assigning an expectation value to every operator in the algebra in that state. A famous construction by \citet{gelfand_imbedding_1943} and \citet{segal_irreducible_1947} (GNS) then secures the existence, for every such state $\psi$, of a Hilbert space $\mathscr{H}_\psi$ with:

\begin{itemize}[label=$-$]
    \item a representation of the algebra $\pi_\psi:\mathcal{A}\to\mathcal{B}(\mathscr{H}_\psi)$, and;
    \item a state $\ket{\psi}\in\mathscr{H}_\psi$ that is cyclic for $\pi_\psi$;\footnote{This means that $\pi_\psi(\mathcal{A})\ket{\psi}$ is dense in $\mathscr{H}_\psi$.}
\end{itemize}
such that $\psi(A) = \bra{\psi}\pi_\psi(A)\ket{\psi}$ for all operators $A\in\mathcal{A}$. Here $\mathcal{B}(\mathscr{H}_\psi)$ denotes the set of bounded operators on $\mathscr{H}_\psi$. The GNS-construction thus allows us to translate from AQFT into the familiar Hilbert space approach. 

If the algebra is equipped with a vector space norm (with $\norm{AB}\leq\norm{A}\norm{B}$) and is also closed in this norm, the algebra is called a \textit{Banach algebra}. If we additionally require that $\norm{A^* A}=\norm{A}^2$, the algebra is a \textit{C*-algebra}. Early pioneers in AQFT, such as \citet{segal_mathematical_1963} and \citet{haag_algebraic_1964}, argued that all physically relevant information is contained in the C*-algebra and that the choice of representation is only a matter of convenience. However, nowadays this position is not considered viable, because there are generally physically interesting operators that are missing from the C*-algebra, for example the stress-energy tensor. For this reason, \citet{wald_quantum_1994}, amongst others, has advocated for a more moderate approach in which C*-algebra or *-algebra (as opposed to the Hilbert space formalism or von Neumann algebra, as will be discussed below) is interpreted rather as a minimal set of observables needed to formulate the structure of the theory, and to which we can add operators that we may find physically interesting. Also, \citet{hollands_axiomatic_2010} have argued that while the additional C*-algebra structure can be useful, it is not necessary for most purposes and therefore preferable to work with the more general *-algebra to avoid arbitrariness. In that case, we work with an algebra of unbounded operators and the relevant relations between the fields are described by the operator product expansion. 

\subsection{Von Neumann Algebras}
\label{section:vN}

Recently, the focus in AQFT has shifted to the study of von Neumann algebras, which will be relevant in the rest of this paper, since in the physically relevant representations the local algebras of observables are von Neumann algebras, and modular-theoretic tools used to analyse horizons and entropy are standardly formulated in that setting.
A von Neumann algebra $\vN$ is a special type of C*-algebra that satisfies the following condition:
\begin{equation}
\label{eq:doublecommutant}
    \vN = (\vN) ''
\end{equation}
where the commutant, $\mathcal{A}'$, of (a Hilbert space representation of) an algebra $\mathcal{A}$, is defined as the set of all operators in $\mathcal{B}(\mathscr{H})$ that commute with all operators in $\mathcal{A}\subset \mathcal{B}(\mathscr{H})$, that is, $\mathcal{A}'= \{B \in\mathcal{B}(\mathscr{H}) | [A,B]=0, \forall A\in \mathcal{A}\}$. If the algebra is unital, von Neumann's Double Commutant Theorem applies, which says that the condition in Eq.~(\ref{eq:doublecommutant}) is equivalent to the condition that the algebra is closed either in the weak or the strong operator topology \citep[see e.g.][Thm. C.127]{landsman_foundations_2017}.

Clearly, since the commutant is defined only for a Hilbert space representation of the algebra, the definition of a von Neumann algebra \textit{does} require us to interpret it as an algebra of operators on a specific Hilbert space.\footnote{There is an abstract counterpart of the von Neumann algebra, called the W*-algebra, that is not defined as an algebra of Hilbert space operators. Instead, a W*-algebra is defined as a C*-algebra that is also the dual space of a Banach space over $\mathbb{C}$. However, this is equivalent to restricting to a particular folium of states on the C*-algebra, and thus to a particular quasi-equivalence class of representations \citep[see][]{haag_local_1996}. The GNS-representations of the W*-algebra all give a von Neumann algebra on the corresponding Hilbert space. And since the representations are quasi-equivalent, these von Neumann algebras are all isomorphic.} So a von Neumann algebra is not representation-independent. There is therefore a trade-off of virtues here between the generality of the *-algebra and the technical power of von Neumann algebras. The use of von Neumann algebras is avoided in the axiomatic approach of \citet{hollands_axiomatic_2010}, which seeks to generalize QFT to arbitrary spacetime backgrounds. By contrast, for a fixed background and an explicit choice of time in a (semi)classical gravitational setting, it is natural to work with von Neumann algebras, and this can be technically very fruitful. In this paper, we will mostly be interested in the case where we adopt the perspective of a particular observer without making assumptions about the background, and we will have to investigate to what extent von Neumann algebras are useful in this setting. 

What makes a von Neumann algebra interesting compared to a more general, representation-independent C*-algebra $\mathcal{A}$ is that it contains a more ``complete'' set of the operators on a Hilbert space. As noted above, we can interpret the C*-algebra as a minimal set of observables needed to formulate the theory. On the other hand, because a von Neumann algebra is closed in the weak operator topology, it contains all operators $\tilde{A}\in\mathcal{B}(\mathscr{H})$ for which:
\begin{equation}
\label{eq:weak}
    |\braket{\psi |\tilde{A}|\psi} - \braket{\psi| A |\psi }| < \varepsilon, \quad A\in\mathcal{A}
\end{equation}
for any $\ket{\psi}\in\mathscr{H}$ and any $\varepsilon>0$. This can be interpreted as saying that for any measurement with finite accuracy $\varepsilon$, the von Neumann algebra $\vN$ contains all operators in $\mathcal{B}(\mathscr{H})$ that cannot be distinguished from operators in the more fundamental C*-algebra $\mathcal{A}$. Moreover, the inclusion of these operators allows us to consider so-called \textit{affiliated} operators whose spectral projections lie in $\vN$ --- these include potentially physically interesting unbounded operators, such as the position and momentum operators \citep{clifton_are_2001}. By contrast, this is not possible with the fundamental C*-algebra, because this algebra generally does not contain any non-trivial projections. 

Von Neumann algebras can be conveniently classified into three types.\footnote{Technically the type classification is defined for factors, which are von Neumann algebras for which $\mathcal{A}\cap\mathcal{A}'=\mathds{1}$. However, all von Neumann algebras can be expressed as a direct integral of factors, see \citet{sorce_notes_2024}.} The type of the algebra tells us 
which familiar concepts of quantum mechanics are present in the algebra \citep{witten_gravity_2022}: Type I is the most familiar and contains, among other things,  pure states, density matrices, and entropies. Type II algebras are more exotic; they have no notion of a pure state, but still allow for density matrices and entropies.  Finally, Type III algebras do not even have density matrices or a notion of entropy.\footnote{Another way to see this is that Type II and Type III algebras have an infinite amount of entanglement built into their structure, as opposed to Type I algebras where entanglement is only a property of states. See \citet{witten_why_2022}.} It is this last type that we typically encounter in QFT. More precisely, the types can be defined as follows:\footnote{See also \citet{landsman_foundations_2017} for a more formal introduction.}

\begin{itemize}
    \item \textbf{Type I} algebras have an irreducible representation on a Hilbert space. This is in fact the more familiar type encountered in quantum theory, the algebra of all bounded operators on a Hilbert space: $\vN=\mathcal{B}(\mathscr{H})$.
    \item \textbf{Type II} algebras do not have an irreducible representation on a Hilbert space. This means that its commutant in any Hilbert space representation is non-trivial: it is another algebra of Type II. Type II algebras do have a notion of trace.\footnote{The trace is defined as a function $\Tr:\mathcal{A}^{\text{vN}}_+ \to [0,\infty]$ on positive operators in $\vN$ (i.e.~operators of the form $A=B^*B$)  with the property $\Tr(AB)=\Tr(BA) \quad \forall A,B\in \mathcal{A}^{\text{vN}}_+$.} If the trace is finite for all elements in the algebra, in which case we can normalize it as $\Tr(\mathbf{1})=1$, the algebra is called Type II$_1$. If the trace cannot be normalized, it is Type II$_\infty$.
    \item \textbf{Type III} algebras do not have an irreducible representation nor do they have a trace. In fact, a Type III algebra only has one representation up to isomorphism, and its commutant in any representation is thus (anti-)isomorphic to itself. 
\end{itemize}

Since the smeared fields are local operators (see footnote \ref{note:smeared}), it makes sense to look at the algebra associated to some particular region of the spacetime manifold (here and in the rest of the paper we will always assume global hyperbolicity\footnote{A globally hyperbolic  manifold has topology $\mathbb{R}\times\Sigma$, where $\Sigma$ is an arbitrary submanifold. This ensures that the manifold can be foliated in a one-parameter family of Cauchy surfaces $\Sigma_t$ with $t\in\mathbb{R}$ that any causal curve can cross only once. Global hyperbolicity is therefore usually taken as a physical consistency requirement to avoid causal pathologies such as closed timelike loops \citep[see][§4.1]{wald_quantum_1994}.}). Such an algebra can be interpreted as representing the observations that can be made within that specific region. The von Neumann algebra associated to a pure state on a complete Cauchy surface is, by virtue of causality, equal to the algebra associated to the entire manifold.\footnote{This is known as the Time-Slice Axiom, originally formulated by \citet{haag_postulates_1962}.} This algebra contains the full set of the observables of the theory, which means that it can be represented as the set of \textit{all} bounded operators on some Hilbert space, and is thus of Type I. On the other hand, the algebra of observables associated to some arbitrary open region that is not a complete Cauchy surface is typically of Type III \citep[see][§V.6]{haag_local_1996}. For example, \citet{leutheusser_causal_2023,leutheusser_emergent_2023,leutheusser_subregion-subalgebra_2025} showed how a Type III algebra can arise from a Type I algebra in a CFT dual to an AdS-Schwarzschild geometry in the semiclassical limit, with the appearance of a sharp horizon which bounds the subregion to which the Type III algebra is dual. Similarly, the algebra of observables associated to the static patch in de Sitter space, which is bounded by a cosmological horizon, is of Type III. 

Type II algebras can arise by taking the `crossed product' of a Type III algebra with its modular automorphism group, which we will define below. To explain this construction, we first briefly review some aspects of Tomita-Takesaki theory \citep{takesaki_tomitas_1970}. A central theorem in this theory is that a cyclic and separating\footnote{A vector is separating if $A\ket{\psi}=0\implies A=0$.} vector $\ket{\psi}$ in a Hilbert space $\mathscr{H}$ on which a von Neumann algebra $\vN$ acts, defines a one-parameter group of automorphisms of the algebra, which is called the modular automorphism group or modular flow. Starting with the the Tomita operator $S_\psi$ defined by  $S_\psi A\ket{\psi} = A^\dagger\ket{\psi}$, the `modular' operator is constructed as $\Delta_\psi=S_\psi^\dagger S_\psi$ and defines a one-parameter group of automorphisms whose action on $\vN$ is given by:
\begin{equation}
\label{eq:modlow}
    \alpha_s^{\psi}(A)= \Delta_\psi^{-is}A\Delta_\psi^{is}\in \vN, \quad\quad \forall \; A \in \vN, \; s\in\mathbb{R}\,.
\end{equation}
The action of this group can be written in a more familiar form as:
\begin{equation}
    \alpha_s(A) = e^{is H_\psi}Ae^{-is H_\psi},
\end{equation}
where $H_\psi=-\log\Delta_\psi$ is called the modular Hamiltonian. 

For a Type III algebra, the modular flow is an outer automorphism for some values of $s$, which means that it is not generated by an element within the algebra itself. Namely, if the automorphism were always inner, this would allow us to define a trace on the algebra, in which case it must be of Type I or II \citep[Thm VIII.3.14]{takesaki_theory_2003}. The converse is also true, so that the outerness of the modular automorphism group can be taken as an alternative defining property of a Type III algebra. On the other hand, for a Type II algebra the modular flow is always an inner automorphism. For a Type I algebra, all automorphisms must be inner \citep[Thm XI.3.7]{takesaki_theory_2003}.

Another important fact that we will use later in Sections \ref{sec:gravdressing} and \ref{section:dS} is that a state $\psi$ which defines a modular flow on the algebra satisfies the Kubo-Martin-Schwinger (KMS) condition. This condition is defined as:
\begin{equation}
\label{eq:KMS}
\mel{\psi}{A\alpha_{i\beta}(B)}{\psi}=\mel{\psi}{BA}{\psi}, \quad \forall A,B\in\vN.
\end{equation}
And $\ket{\psi}$ satisfies \eqref{eq:KMS} with respect to its own modular flow $\alpha_t^{\psi}$ as defined by \eqref{eq:modlow} at an inverse temperature $\beta=1$ \citep{takesaki_tomitas_1970}. The KMS condition can be seen as a generalization of the Gibbs condition $\psi = Ne^{-\beta H}$ to systems with an infinite number of degrees of freedom \citep[see e.g.][]{haag_equilibrium_1967}. In other words, the state can be interpreted as being thermal with respect to the modular flow that it defines on the algebra. 

The modular automorphism group allows us to construct a Type II algebra from a Type III algebra via the crossed product. This is relevant because Type II algebras contain a notion of density matrices and entropies, and thus enable us to do more physically interesting calculations. The crossed product of $\vN$ by the additive group $\mathbb{R}$, denoted $\vN \rtimes \mathbb{R}$, is constructed with the help of an auxiliary Hilbert space $L^2(\mathbb{R})$ of square-integrable functions on $\mathbb{R}$. Paradigmatically, $L^2(\mathbb{R})$ is the Hilbert space isomorphic to that of a particle moving on a line, with canonical position and momentum operators $q$ and $p$.  The result of the crossed product is a new algebra that acts on the Hilbert space $\mathscr{H}\otimes L^2(\mathbb{R)}$, where $\mathscr{H}$ is the Hilbert space on which the original algebra $\vN$ acted. A vector $\ket{\psi}\in\mathscr{H}\otimes L^2(\mathbb{R})$ can be written in a momentum basis $\ket{\psi}=\int\mbox{d}p \ket{\psi(p)}\otimes \ket{p}$ and as such be seen as a function $p\mapsto \ket{\psi(p)}$ \citep{jensen_generalized_2023}. Consider again a state $\psi$ on $\vN$ that defines a one-parameter group of automorphisms as in Eq.~(\ref{eq:modlow}). Then the {\it crossed product}
 of $\vN$ and $\mathbb{R}$ with respect to the vector $\psi$, i.e.~$\vN \rtimes_\psi \mathbb{R}$, is the smallest von Neumann algebra containing operators of the form $A\otimes \mathbf{1}$ and $e^{is H_\psi} \otimes e^{is q}$, with $A\in \vN$ and $s\in\mathbb{R}$. Intuitively, this latter operator can be understood as simultaneously enacting a global `time' shift on $\mathscr{H}$ by $e^{is H_\psi}$ and a corresponding shift $\ket{p}\to \ket{p+s}$ in the auxiliary Hilbert space $L^2(\mathbb{R})$. The `momentum' $p$ in the auxiliary Hilbert space thus keeps track of the elapsed time in $\mathscr{H}$. It can be checked that the operators $A\otimes \mathbf{1}$ and $e^{is H_\psi} \otimes e^{is q}$, additively written as $Ae^{is H_\psi}\otimes e^{is q}$, indeed form an algebra \citep{witten_gravity_2022}. We can denote 
it as:
\begin{equation}
\label{eq:cross}
    \vN \rtimes_\psi \mathbb{R} = \{A\otimes 1,\; e^{is H_\psi} \otimes e^{is q}| \;A\in\vN; \; s\in\mathbb{R} \}''.
\end{equation}

What is important here is that the crossed product can change the Type of the algebra. If the original algebra $\vN$ is of Type III, then the crossed product with its modular automorphism group produces a von Neumann algebra of Type II \citep{takesaki_duality_1973}. To see this, consider the fact that the modular automorphism group of the original algebra naturally extends to an automorphism of the crossed product algebra via $\alpha_t \otimes \mathbf{1}$, where $\alpha_t$ is the action of the group on $\vN$ defined in Eq.~(\ref{eq:modlow}). This extension is the modular automorphism group of the dual weight $\hat\psi$ of the state $\psi$ on $\vN \rtimes_\psi\mathbb{R}$. However, the modular flow is now an inner automorphism, as it can be represented by a unitary operator $e^{isH_\psi}\otimes e^{isq}$ which is included in the crossed product algebra itself. This means that we can define a trace on $\vN \rtimes_\psi \mathbb{R}$ via $\Tr(A) = \hat\psi(e^{-H_\psi-q}A)$. Therefore, the crossed product algebra of a Type III algebra with its modular automorphism group is of Type II.

Interestingly, and perhaps surprisingly, the resulting algebra $\vN \rtimes \mathbb{R}_\psi$ does \textit{not} depend on the state $\psi$ that is used to construct it.\footnote{That is, the crossed products constructed using two different states are not just isomorphic but related by a unique conjugation, which means that they should be interpreted as really being physically the same. This can be proved using Connes' cocycles, see \citet[§3.2]{witten_gravity_2022}.} This mathematical fact will be important because, as we will see in Section \ref{sec:gravdressing}, what CLPW call `gravitationally dressing observables to the worldline of an observer' 
amounts to taking the crossed product of that algebra with the modular automorphism group defined by the clock of the observer, turning the Type III algebra into a Type II. Since in the algebraic setup states are associated to specific backgrounds, the state-independence of the crossed product algebra can be interpreted as a form of background-independence in the sense of (ii) in Section \ref{sec:introduction} \citep{witten_gravity_2022}. We will return to this in Section \ref{sec:newnotion}.

\subsection{Gravitationally Dressing Operators to the Worldline of an Observer}
\label{sec:gravdressing}
In this section, we introduce CLPW's framework. The goal is to formulate an algebra of observables localised along the worldline of an observer in a general, semiclassical spacetime setting, assuming Newton's constant $G$ to be small compared to the characteristic de Broglie wavelength of the matter fields, such that gravity is weakly coupled to matter and can be treated perturbatively. Following the arguments made in the introduction, we will explicitly include the observer as a quantum degree of freedom in the construction, and require that the algebra is independent of the state of both the spacetime and matter fields and the state of the observer itself. 

The setup follows \citet{witten_background_2025}, and consists of a worldline $\gamma:\mathbb{R}\to M$, describing the trajectory of an observer travelling along a geodesic on a manifold $M$. We focus on the algebra of matter fields that the observer can measure along $\gamma$ and we make no assumptions about the background manifold except that it is globally hyperbolic.
For simplicity, we restrict our attention to a single scalar field $\phi(x)$ that the observer can measure along its worldline $\gamma$. Let $\tau\in \mathbb{R}$ be the observer's proper time that parametrises $\gamma$, then $\phi(\tau)\equiv \phi(\gamma(\tau))$ denotes the field at that point along $\gamma$. These will be the building blocks used to construct our algebra including the observer. 

However, since only differences in proper time between two spacetime points have physical significance, the fields should be invariant under reparametrisations of the form:
\begin{equation}
\label{eq:repara}
    \tau\to\tau+c
\end{equation}
which can be seen as a relabelling that changes the zero-point of $\tau$. In the algebraic approach, we would like to impose the constraints at the level of the algebra generated by the fields $\phi(\tau)$.\footnote{See e.g.~\citet[§3]{jensen_generalized_2023}, for a review of gravitational constraints in algebraic quantisation.} We denote the operator enacting the transformation \eqref{eq:repara} by $H_\gamma$, which thus acts on the fields as:
\begin{equation}
    \label{eq:hmc}
    [H_\gamma,\phi(\tau)] = -i\partial_\tau\phi(\tau)\,.
\end{equation}
where we put in the $i$ for later convenience. In the presence of gravity, we would ideally like to interpret the operator $H_\gamma$ on the algebra as the generator of a diffeomorphism of the background that maps $\gamma$ to itself and shifts $\tau$ by a constant, thereby giving $H_\gamma$ a direct geometric interpretation in the background. But it turns out that this can only be the case if it generates a (conformal) symmetry of the background \citep{sorce_analyticity_2024}. More generally, using the algebraic language introduced in Section \ref{section:vN}, $H_\gamma$ can be seen as the generator of a modular flow on the algebra, that typically corresponds to a non-local operator.\footnote{ The state corresponding to this modular flow as in Eq.~\eqref{eq:modlow} is not unique: the modular flows generated by different states are all related by an inner automorphism, as captured by Connes' Cocycle Radon-Nikodym theorem \citep[see][]{connes_von_1994}.} In any case, we take Eq.~\eqref{eq:hmc} to be the definition of $H_\gamma$.\\
Reparametrisation invariance now requires that we impose $H_\gamma$ as a constraint, setting Eq.~\eqref{eq:hmc} to zero, such that $\phi(\tau)$ is now constant along $\gamma$. Clearly, this does not lead to the desired result: there is no dynamics at all. This is a particular manifestation of the problem of time in quantum gravity: there is no dynamics if the Hamiltonian is imposed as a constraint \citep[see e.g.][]{butterfield_spacetime_1999,callender_physics_2001,kuchar_time_2011}. As originally proposed by \citet{page_evolution_1983}, in order to have dynamics in quantum gravity we need to relate the observables to an internal degree of freedom. So far, we have only considered matter fields along some worldline and not included the observer as a quantum system. If we do this, the observer's system can model a clock. This is consistent with the proposed solution to problem (iii) in Section \ref{sec:introduction}: we need to include an internal observer to have an interpretation of the algebra in terms of observations. 

This is indeed the approach that CLPW take. To obtain a non-trivial algebra, they propose to include an additional degree of freedom that represents the observer. This additional degree of freedom can model an internal clock with respect to which the dynamics of the observables is defined. More concretely, the observer is modelled by a point particle moving along the geodesic, with canonical variables $\omega$ and $t$, and described by the action: 
\begin{equation}
\label{eq:obs-action}
    I_\text{obs} = \int_\gamma d\tau \left(t\dot{\omega}-\omega\right),
\end{equation}
where the CCR read $[\omega,t]=i$ (in units $\hbar=c=1$). Of course, other models for the observer are possible, but this seems to be the simplest model at hand.

Thus the total system now consists of the matter fields and the observer: the matter fields are represented by the field operators acting along the worldline of the observer, with a Hamiltonian $H_\gamma$ as defined by Eq.~(\ref{eq:hmc}), and the observer is modelled by the action in Eq.~(\ref{eq:obs-action}). So far these are two commuting systems. In that case, the total Hilbert space can be written as a tensor product $\mathscr{H}_{\text{tot}}= \mathscr{H}_\text{QFT}\otimes\mathscr{H}_{\text{obs}}$, where $\mathscr{H}_\text{QFT}$ is the Hilbert space on which the field operators $\phi(\tau)$ act (which depends on a choice for $M$ and $\gamma$) and $\mathscr{H}_{\text{obs}}$ is the Hilbert space associated to the action in Eq.~(\ref{eq:obs-action}). We could also add terms to the observer's action that couple the observer's system to the fields $\phi(\tau)$, allowing for interactions between the observer and the fields that can be interpreted as measurements. However, as we will return to in Section \ref{sec:newnotion}, this will not qualitatively change the algebra \citep[see also][]{witten_background_2025}. So for simplicity, we will not consider such a coupling for now, even though the observer is interpreted to have access to the observables along $\gamma$.

The Hamiltonian corresponding to the action Eq.~(\ref{eq:obs-action}) is:
\begin{equation}
\label{eq:hobs}
    H_\text{obs} = \omega\,.
\end{equation}
Since $\omega$ represents the observer's energy, its canonical conjugate $t$ can be interpreted as reading the observer's time. So it is in this sense that the additional degree of freedom modelling the observer can be interpreted as a simple quantum clock. $H_\text{obs}$ is the generator of shifts of the clock time $t$, as can be seen by taking the standard Schrödinger representation of the CCR: $\omega=i\partial_t$.

Instead of requiring operators to commute with the Hamiltonian $H_\gamma$, we now require them to commute with the combined Hamiltonian $\Ha{tot} = H_\gamma+\Ha{obs}$. Imposing 
\begin{equation}
\label{eq:htc}
    [\Ha{tot},\phi(\tau)]=0\,,
\end{equation}
and combining this with Eq.~(\ref{eq:hmc}), implies $\partial_\tau\phi(\tau)=\partial_t \phi(\tau)$. This is easily satisfied by setting $t=\tau + c$, where $c$ is a constant (cf. Eq.~(\ref{eq:repara})). Thus the Hamiltonian constraint, Eq.~\eqref{eq:htc}, requires us to identify the time $t$ of the internal clock of the observer with the parameter $\tau$, up to some additional constant. This is consistent with the interpretation in GR that the parameter $\tau$ is the proper time along a timelike curve: here, it is the time on the observer's clock along $\gamma$. The algebra of observables $\A{obs}$ along the worldline can now be generated by the family of operators:
\begin{equation}
    \phi_c(t)\equiv\phi(t+c)\,.
\end{equation}
Additionally, the operator $H_{\text{obs}} = \omega$ evidently commutes with $H_{\text{tot}}$, and should be included in the algebra. So $\A{obs}$ is now defined as the smallest algebra containing the operators $\phi_c(t)$ and bounded functions of $H_{\text{obs}}$. 

For a particular choice of $M$ and $\gamma$,  we can interpret $\A{obs}$ as an algebra of operators acting on the Hilbert space $\mathscr{H}_{\text{tot}}=\mathscr{H}_\text{QFT}\otimes L^2(\mathbb{R})$, and close $\A{obs}$ to a von Neumann algebra. However, without making such a choice, we can interpret $\A{obs}$ simply as an operator product algebra or *-algebra, which is then viewed as the underlying structure of the theory.\footnote{What Witten seems to have in mind is only an operator product algebra, so with even less structure and hence more general than a *-algebra. However, the relation between an operator product algebra and a von Neumann algebra is less clear, and it seems that we need to assume *-algebra structure before the algebra can be closed in a particular representation. We will, for simplicity, restrict to the case where the field operators actually form a *-algebra, which is widely discussed in the literature. In any case, for our analysis of background-independence it does not really matter whether $\A{obs}$ is an operator product algebra of *-algebra, since the relevant point in our argument is that the algebra is independent of a particular state or representation, which is true for both.\label{note:opa}} 

In the case where $H_\gamma$ generates a modular flow on $\A{}$, such that $\phi_c(t) = e^{itH_\gamma}\,\phi_c(0)\,e^{-itH_\gamma}$, the algebra can equivalently be written as:
\begin{equation}
\label{eq:Aobs2}
    \A{obs} = \{ \phi_c\otimes\mathbf{1},\;e^{i s H_\gamma}\otimes e^{is H_{\text{obs}}}|\;s\in\mathbb{R}\}\,,
\end{equation}
which, comparing to Eq.~\eqref{eq:cross}, shows that $\A{obs}$ is the crossed product of $\A{}$ with the modular automorphism group defined by $H_\gamma$. This implies that, since $\A{}$ is generally of Type III, $\A{obs}$ is of Type II and has a notion of denisty matrices and traces. However, $H_\gamma$ is usually not proportional to the modular Hamiltonian of some particular state, which is typically a non-local object, except when it is a symmetry of the local region. This is the case for example for a boost in Rindler space, or a time shift of the static patch in dS, which will be discussed in Section \ref{section:dS}. More generally, $H_\gamma$ can be approximated as a modular Hamiltonian if it approximately acts as a boost near an entangling surface \citep[see][§4]{jensen_generalized_2023}. As we will also discuss in the next section, \citet{witten_background_2025} conjectures that, in any spacetime, $\A{obs}$ will be of Type II. 

To get a bit more intuitive understanding of the physical significance of the crossed product algebra $\A{obs}$, we may note, following \citet{de_vuyst_gravitational_2024}, that CLPW's construction is essentially the same as the \citet{page_evolution_1983} mechanism. This mechanism shows that even though closed quantum-gravitational systems have to be static from a global (outside) perspective because the symmetries have to be imposed as constraints, observables can \textit{appear} dynamical to an observer within the system itself, who relates them to some internal clock variable. In this case, the role of the clock is played by an auxiliary Hilbert space $L^2(\mathbb{R})$ (which is the same setup as Page and Wootters considered), which acts as a \textit{quantum reference frame} \citep[see e.g.][]{hohn_trinity_2021}. It can be seen that by writing:
\begin{equation}
\label{eq:schrödingerequation}
    [H_\gamma,\phi(t)]=i\partial_t\phi(t)\,,
\end{equation}
we recover the usual Schrödinger equation in the Heisenberg picture. So we expect the observer to see the usual dynamics for the fields with respect to their own internal clock system, even though, as required by Eq.~(\ref{eq:htc}), the fields must be static from a global point of view. 

Finally, for a more realistic scenario, we may require that the observer's energy is bounded from below. This can be achieved by adding a rest mass $m$ to the observer's Hamiltonian in Eq.~\eqref{eq:hobs}, and conjugate the algebra with the Heaviside operator $\Theta(\omega)$ that projects onto states with $\omega \geq0$:
\begin{equation}
    \hat{\phi}_c(t)\equiv\Theta(\omega)\phi_c(t)\Theta(\omega)\,.
\end{equation}
If $\omega$ is bounded from below, its canonically conjugate $t$ can no longer be defined as a self-adjoint operator, and thus our quantum mechanical clock cannot keep track of the time perfectly but will be fuzzy. However, CLPW circumvent this by arguing that the clock is still able to `tell time very well, for a long period' --- namely, for states with support away from $\omega =0$, which are the relevant states at least in the de Sitter context.\footnote{Alternatively, one can define the time observable as a positive operator-valued measure \citep[see e.g.][]{hohn_equivalence_2021}.} With the spectrum bounded from below, in cases where the spacetime region accessible to the observer is bounded by a horizon, the algebra is expected to become a Type II$_1$ von Neumann algebra in representations where the state is normalizable in the corresponding spacetime, and Type II$_\infty$ otherwise,  \citep{witten_background_2025}. The difference is that in a Type II$_1$ algebra the trace is normalizable, while in a Type II$_\infty$ it is not. 

However, as pointed out by \citet{kolchmeyer_chaos_2024}, we have to be careful if we want to include any sort of coupling between the observer and the matter fields, or include backreaction of the observer and fields on the metric. In de Sitter-type spacetimes, the expansion of the universe can let a small perturbation of the observer's worldline grow to arbitrary sizes. If the mass of the observer is large with respect to the energy of the interaction, the time needed to change the trajectory of the observer by a appreciable amount is of the order of the scrambling time, which goes as the logarithm of the observer's mass. Therefore, when including such couplings, the observer can be approximated to follow a classical worldline only if we assume that the length of the their worldline is much shorter than the scrambling time or if we take the limit $m\to\infty$.\footnote{But note that in this latter case, the order of limits is important: we first have to take the limit $G\to0$, and then $m\to \infty$, to prevent the observer from backreacting on the metric and forming a black hole.} If this is not the case, the observer's worldline becomes subject to quantum uncertainty. This means that, if the worldline extends from past to future infinity, the observer would be able to explore an entire Cauchy surface by acting on the fields and thus diverting their worldline, in which case the algebra becomes of Type I, as Kolchmeyer and Liu show. A complementary perspective is provided by \citet{chen_algebra_2026}, who introduce what they call a covariant observer in de Sitter space. Instead of modelling the observer as a fully quantised relativistic particle, as in the setup of Kolchmeyer and Liu, whose trajectory is uncertain and can change due to recoil, they model the observer as an ideal quantum reference frame supporting a superposition of classical trajectories, for each of which the observer is restricted to its own static patch. As a result, after imposing the de Sitter constraints they do get a Type II algebra, which they interpret as an `averaged' modular crossed product algebra. 

Let us quickly summarize what we have discussed so far. We started with a generic worldline $\gamma$ in a background $M$ and considered a set of matter field operators $\phi(\tau)$ along this worldline. To make these operators physically significant, we needed to impose a  constraint that ensured that the algebra is invariant under the symmetries of spacetime. However, using the generator of diffeomorphisms by itself led to a trivial result. Instead, we needed to augment the system with an auxiliary Hilbert space that behaves as a clock, and is interpreted as a minimal model for an observer (or a quantum reference frame). We arrived at an algebra of observables $\A{obs}$ along $\gamma$. From the point of view of the observer, $\A{obs}$ describes the usual Schrödinger dynamics with respect to their own internal clock system. Different states on the algebra can be associated to different choices of background, and when closed in a particular representation, the algebra is expected to be generally a Type II von Neumann algebra. 

\subsection{An Algebra for de Sitter Space}
\label{section:dS}

In the previous section, our minimal model for the observer and the matter field did not use the details of the spacetime background. In this Section, we consider the case of a scalar field $\phi$ in a de Sitter background, that is associated to a specific state on the algebra. Namely, for such a QFT there is a natural vacuum state that is invariant under the $SO(D,1)$ isometry, usually called the Bunch-Davies state, that we call $\Psi_{\text{dS}}$ \citep{greene_universal_2006}.\footnote{This vacuum is actually part of a family of vacua parametrized by $\alpha$, where the Bunch-Davies vacuum corresponds to $\alpha=-\infty$.} This state has the property that it satisfies the Kubo-Martin-Schwinger condition:
\begin{equation}
\label{KMS}
   \braket{\Psi_{\text{dS}}|\phi_{c}(t)\phi_{c'}(t')|\Psi_{\text{dS}}}=  \braket{\Psi_{\text{dS}}|\phi_{c'}(t'-i\beta_{\text{dS}})\phi_c(t)|\Psi_{\text{dS}}},
\end{equation}
with respect to the time measured along any timelike geodesic \citep{witten_background_2025}. This can be interpreted as saying that the state looks thermal to any such observer with an inverse temperature $\beta_\text{dS}=2\pi r_{\text{dS}}$ \citep[see e.g.][]{haag_equilibrium_1967}.  

We now want, as in the previous section, to explicitly include the observer into our description. The region that is causally accessible to a geodesic observer in de Sitter space is called the static patch; this name comes from the fact that the static patch has a Killing vector field that is everywhere timelike and becomes null on its boundary, which forms the observer's cosmological horizon \citep{gibbons_cosmological_1977}. Along the observer's worldline, this vector field acts as the generator of shifts of the observer's proper time, just like $H_\gamma$ in Section \ref{sec:gravdressing}. 

We define an algebra of observables $\A{obs}$ following the procedure outlined in Section \ref{sec:gravdressing}, i.e.~gravitationally dressing the fields to the worldline of the observer \citep[for more details, see][§2.4]{chandrasekaran_algebra_2023}.  As we noted there, a specific choice of background should allow us to define a state on the algebra that is KMS with respect to the time flow on $\A{obs}$, i.e.~the time flow along the worldline $\gamma$. An obvious candidate to perform this role is the Bunch-Davies state $\Psi_\text{dS}$, suitably modified to include the observer.  Since $\Psi_{\text{dS}}$ is interpreted as a highest entropy state for de Sitter space without an observer, it makes sense to 
augment it by adding a thermal distribution for the observer's energy,
which is written as:
\begin{equation}
    \Psi_\text{dS+obs} = \sqrt{\beta_{\text{dS}}}\;e^{-\beta_{\text{dS}}\,\omega/2}\Psi_{\text{dS}}\,,
\end{equation}
 interpreted as a state on the algebra $\A{obs}$. Then, we can use $\Psi_\text{dS+obs}$ to construct a Hilbert space $\mathscr{H}_{\Psi_\text{dS+obs}}$ representation for the algebra via the GNS-construction. This will be the Hilbert space $\mathscr{H}_{\Psi_\text{dS+obs}}= \mathscr{H}_\text{dS}\otimes\mathscr{H}_\text{obs}$, where $\mathscr{H}_\text{dS}$ is the Hilbert space of the empty de Sitter vacuum, i.e.~without an observer \citep{chandrasekaran_algebra_2023}. 
 
The KMS condition Eq.~(\ref{KMS}) now implies not only that $\Psi_{\text{dS+obs}}$ is KMS, but even that it is tracial: 
\begin{equation}
    \bra{\Psi_\text{dS+obs}}\phi_c(t)\phi_{c'}(t')\ket{\Psi_\text{dS+obs}}=\bra{\Psi_\text{dS+obs}}\phi_{c'}(t')\phi_c(t)\ket{\Psi_\text{dS+obs}},
\end{equation}
where $\phi_c(t),\; \phi_{c'}(t')\in \A{obs}$, the proof of which is given in two different versions by \citet{witten_algebras_2024,witten_background_2025}. This means $\Psi_\text{dS+obs}$ can be used to define a trace:
\begin{equation}
    \Tr(\phi_c(t)) \equiv \braket{\Psi_\text{dS+obs}|\phi_c(t)|\Psi_\text{dS+obs}}, \quad \phi_c(t)\in\A{obs}\,,
\end{equation}
which is in agreement with our earlier prediction that $\A{obs}$ is a Type II von Neumann algebra.  If the observer's energy spectrum is required to be bounded from below,\footnote{This can be imposed by conjugating all operators $\phi_c(t)\in\A{obs}$ by the Heaviside projection operator $\Theta(\omega)$, see \citet{chandrasekaran_algebra_2023}} the trace of the identity element:
\begin{equation}
    \Tr(\mathbf{1}) = \beta_\text{dS}\int_0^\infty d\omega \; e^{-\beta_\text{dS}\omega} |\Psi_\text{dS}|^2\,,
\end{equation}
is finite and in particular can be set to one if $\Psi_\text{dS}$ is suitably normalized, which defines our algebra $\A{obs}$ to be of Type II$_1$ \citep{chandrasekaran_algebra_2023}. If $\omega$ is not bounded, the trace of the identity is divergent and the algebra is of Type II$_\infty$. 

The GNS-construction ensures that the state $\Psi_\text{dS+obs}$ is cyclic for the representation of the algebra on $\mathscr{H}_{\Psi_\text{dS+obs}}$. This means that states $\A{obs}\ket{\Psi_\text{dS+obs}}$ form a dense set in $\mathscr{H}_{\Psi_\text{dS+obs}}$, i.e.~any state in $\mathscr{H}_{\Psi_\text{dS+obs}}$ is arbitrarily close to a state in $\{ A\ket{\Psi_\text{dS+obs}}\ | A\in\A{obs}\}$. States of this form can be represented by a density matrix $\rho_A = AA^\dagger$, and we can associate a von Neumann entropy to them:
\begin{equation}
\label{eq:vNS}
    S(\rho) = -\Tr(\rho \log\rho)\,.
\end{equation}
It is then easy to see that the state $\Psi_\text{dS+obs}$, for which $\rho =\mathbf{1}$, has zero entropy. And it can be proven that the entropy of any other state is strictly smaller than zero \citep{witten_background_2025}.\footnote{The fact that the entropies are negative is not alarming. A type II algebra can be seen as a system of $N$ entangled qubit pairs where we take $N \to \infty$. The von Neumann entropy of the maximally entangled state scales with $N$ and thus diverges. This means that the entropy defined in a Type II algebra should be seen as a ``renormalized'' von Neumann entropy where we subtract the entropy of the maximally entangled state, so the maximum entropy is set to zero. See \citet{witten_why_2022}} So $\Psi_\text{dS+obs}$ is indeed the state of maximum entropy, as it was constructed. How should this entropy be interpreted? Since de Sitter space is expanding, any field excitations present in the static patch will exit this region through the future horizon within a finite time. The second law of thermodynamics thus implies that for de Sitter space, without the inclusion of an observer, the empty static patch is a state of maximum entropy.\footnote{This argument was originally formulated by \citet{bousso_positive_2000}.} Now that we have included an observer---which, by definition, cannot exit the static patch---we expect the system to evolve to a state in which the observer is in thermal equilibrium with an otherwise empty static patch at the de Sitter temperature, which is precisely what $\Psi_{\text{dS+obs}}$ represents \citep{witten_background_2025}. \citet[][ § 3]{chandrasekaran_algebra_2023} show that the entropy defined in Eq.~(\ref{eq:vNS}) agrees (up to a renormalization constant) with the generalized entropy associated to the cosmological horizon:
\begin{equation}
\label{eq:Sgen}
    S_{\text{gen}} = \frac{A_H}{4G}+S_{\text{out}}\,,
\end{equation}
where $A_H$ is the area of the horizon and $S_{\text{out}}$ the entropy of the fields outside of the horizon (i.e.~inside the static patch), which for late times corresponds to the entropy of the observer. 

Note that, since states $\A{obs}\ket{\Psi_\text{dS+obs}}$ are dense in $\mathscr{H}_{\Psi_\text{dS+obs}}$, the GNS Hilbert space describes $\mathcal{O}(1)$ perturbations around the input state \citep{witten_background_2025}. On the other hand, $\Order{1/G}$ perturbations are associated to different Hilbert space representations. This can also be seen from Eq.~(\ref{eq:Sgen}), which shows that $\Order{1/G}$ perturbations affect the total entropy in the system, that is related to the size of the Hilbert space.

\subsection{Comments on Generalisation to Arbitrary Spacetimes}\label{section:gen}
The example of a de Sitter space background in the previous section confirms the earlier analysis that gravitationally dressing the observables to the worldline of an observer produces a Type II algebra rather than the Type III algebra one usually encounters in QFT, including a notion of entropy that agrees with what we would expect from different considerations. To what extent these results generalize to arbitrary spacetime backgrounds, remains an open question.  \citet{witten_background_2025} conjectures that in a general spacetime the `no-boundary' state introduced by \citet{hartle_wave_1983}, suitably modified to include the observer, plays a similar role as the Bunch-Davies state in de Sitter and can be defined as a tracial state on the algebra. The Hartle-Hawking state of a background $\langle M, \phi_M \rangle $ --- consisting of a manifold $M$ and a set of field configurations $\phi_M$ on that manifold (see also Secion \ref{section_definitions}) --- can be obtained by an Euclidean path integral of the form:
\begin{equation}
    \phi_{HH}(M,\phi_M)=\sum_{\partial N=M}\int_{\eval{\phi}_{\partial N}=\phi_M} D\phi \;e^{-I(\phi)}
\end{equation}
where the sum is over all topologies $N$ that have $M$ as their only boundary, and $I$ is the Euclidean action, which includes the metric as a dynamical field. Some evidence for the idea that for a general background $\phi_{HH}$ plays a similar role as the Bunch-Davies state in the de Sitter case is that the Hartle-Hawking state reduces to the Bunch-Davies state in the limit $G\to0$ in the case of a positive cosmological constant.

At the same time, it should be emphasised that the de Sitter case is special and that it is not immediately clear how all aspects of the construction in Section \ref{section:dS} carry over to a more general setting. The static patch of de Sitter space has an everywhere timelike Killing vector field that is conveniently identified with the clock along the observer's worldline. According to the procedure set out in Section \ref{sec:gravdressing}, we can always define a time along a worldline $\gamma$ as a modular flow on the algebra, but it remains unclear whether this time can be identified with a local geometric flow --- or in other words, whether the modular Hamiltonian $H_\gamma$ corresponds to a vector field and is the generator of a local diffeomorphism. In fact, as we remarked in Section \ref{sec:gravdressing}, this can only be the case if $H_\gamma$ generates a (conformal) symmetry of the background. And thus, because a general spacetime does not admit any symmetries, the modular flow of the worldline generally does not correspond to a local geometric flow, which complicates its interpretation as a local time flow. 

A possible way around this is to appeal to the fact that, even in a general background, spacetime is locally flat. This means that, locally, we can always choose coordinates that admit an approximate Killing vector, such that for any point along the worldline $\gamma$ the modular flow defined by $H_\gamma$ corresponds to a local time flow to a good approximation on length scales that are much smaller than the curvature scale of the background.\footnote{This can be made precise with the use of Fermi normal coordinates, see \citet{manasse_fermi_1963}.} Suppose that we take some small but finite segment of the worldline $l$. Then, according to the timelike tube theorem, the algebra of observables in an infinitesimal tube around this segment is the same as the algebra of the timelike envelope of $l$, which, if $l$ is small enough, approximates a finite causal diamond in Minkowski space that is conformally equivalent to the Rindler wedge \citep{hislop_modular_1982}. In flat space, the modular flow associated to the Minkowski vacuum via Tomita-Takesaki theory on the algebra of the Rindler wedge has a precise geometric meaning: it corresponds to the boost in the wedge.\footnote{This is known as the Bisognano-Wichmann Theorem (\citealt{bisognano_duality_1975,bisognano_duality_1976}; see also \citealt[§V.4]{haag_local_1996})} In a general background, no such vacuum is available, but any globally hyperbolic spacetime admits a large class of Hadamard states that locally behave as the Minkowski vacuum (\citealt{kay_theorems_1991}, see also \citealt[§4.6]{wald_quantum_1994}). It is therefore conceivable that one could identify a Hadamard state whose modular flow approximately acts as the boost around $l$ and that satisfies an approximate KMS condition --- though whether such a state exists and whether it would be sufficient to recover a Type II algebra globally, remain open questions that need further investigation. 

To be sure, if this appeal to local flatness works, it does not mean that we can recover a global modular flow along the entire worldline $\gamma$ by patching the local regions together. The local approximations to $H_\gamma$ effectively give a continuum of approximately defined Hamiltonian constraints that would have to be assembled into a single global constraint of the form Eq.~\eqref{eq:htc}.\footnote{This connects to the problem of the infinite number of Hamiltonian constraints that one encounters when trying to canonically quantise gravity and the effect that all physical observables need to be non-local that has been widely discussed in the physics literature, in particular in the loop quantum gravity and (more recently) the QRFs communities. See e.g.~\citet{goeller_diffeomorphism-invariant_2022} for a recent systematic discussion using relational and dressed observables or \citet{francois_geometric_2025} for a technical presentation of a unified framework for relational observables and dressing methods in gravity and gauge theories.} However, since in a general spacetime with no symmetries the true $H_\gamma$ is non-local, there is no reason to expect that it can be reconstructed by gluing together local contributions defined in Rindler-like patches. In other words, the strategy may provide a local approximation scheme for the time flow seen by the observer that works in a general setting, but it is not expected to yield a satisfactory global picture. 

A more detailed discussion of the difficulties that arise in generalising the analysis of Section \ref{section:dS} is beyond the scope of this paper and will be deferred to future work. Let us therefore repeat that the claim that we present here is simply the following: that the underlying algebraic framework of Section \ref{sec:gravdressing} is completely general; that it gives useful and sensible results in at least one important context: that of de Sitter spacetime; and that we expect that in a general setting similar results can be produced by the framework at least in a local approximation. The significance of an underlying algebraic structure that is completely general, yet in itself incomplete, in connection with idea of background-independence, is the topic of the philosophical discussion to which the remainder of this paper will be dedicated, and will form the basis for the notion of algebraic background-independence introduced in Section \ref{section_pbi}. \\

\section{Notions of Background-Independence}
\label{section_definitions}

Background-independence is widely regarded as a central feature of GR: a feature that is desirable for any theory of quantum gravity. It should be emphasised that background-independence is \textit{not} the same as general covariance, since in principle any theory can be cast into a generally covariant form (as was famously pointed out by \citealt{kretschmann_uber_1917}), nor is diffeomorphism-invariance the same as background-independence (see for example the argument of \citealt{pooley_background_2017}). However, there is no consensus among physicists or philosophers about how to best formulate background-independence, or about how this feature of GR should carry over to future theories.

Given the plurality of definitions, Read argues that background-independence should be seen as a heuristic principle that can guide us in the construction of new theories, rather than as a norm for the justification of existing theories. Others, such as \citet{de_haro_conceptual_2016}, have argued that, while more far-reaching interpretations of background-independence may be taken as a heuristic principle, background-independence in the sense of GR (which they call the `minimalist' sense) should be seen as a minimal consistency requirement for a theory of quantum gravity. This is also a common opinion among physicists.

In this section, we will briefly review three main notions of background-independence in the philosophical literature: absolute object-type definitions, definitions based on variational principles, and Belot's proposal in terms geometric degrees of freedom. We formulate a spacetime theory, according to the standard semantic conception, as a class of models \citep[see e.g.][]{van_fraassen_extension_1970}. We take a model of spacetime to be a pair $\langle M,\phi\rangle$, of a manifold $M$ and a set of fields $\phi$, that satisfy a set of field equations. The fields $\phi$ will generally include the metric field, $\metric$, and matter fields, $\varphi_a$, i.e.~we will write them collectively as follows: $\phi:=\{\metric,\varphi_a\}$. Following \citet{pooley_background_2017} and \citet{read_background_2023}, we will call the space of all such models that make up the theory before imposing the equations of motion the space of \textit{kinematically possible models} (KPMs) $\mathscr{K}$. The subspace $\mathscr{D}\subset \mathscr{K}$ of models that obey the equations of motion is called the space of \textit{dynamically possible models} (DPMs). 

\subsection{Absolute Object-type Definitions}\label{ao}
Absolute object-type definitions can be seen as the earliest approaches to defining background-independence. The idea behind these is that what distinguishes GR from other theories is that it lacks a non-dynamical object, which is the same for all DPMs. Such objects were called `absolute objects' by \citet[see p.~75 for a formal definition]{anderson_principles_1967}.  This suggests that background-independence can be defined as the absence of absolute objects. However, Anderson's original definition of absolute object is perhaps too wide, because arguably also GR itself contains such objects \citep{pitts_absolute_2006}. Therefore, \citet[p.~22]{read_background_2023} proposes a refinement in the form of the narrower notion of absolute fields, which he defines as geometric objects that are explicitly specified in the KPMs of the theory and that are equivalent (up to isomorphisms) in all DPMs of the theory. This excludes pathological examples. Background-independence may be defined as the absence of any absolute fields:
\begin{definition}[Background-independence in terms of absolute fields]
\label{def:BIAF}
A theory is background-independent if and only if it contains no absolute fields. 
\end{definition}
A possible pitfall for Definition \ref{def:BIAF} is that it can seem arbitrary what objects are explicitly specified in the KPMs of the theory \citep[see][p. 23]{read_background_2023}.

In conclusion, absolute object-type definitions are based on the intuition from GR that even spacetime itself is dynamical. They generalise this intuition by defining background-independence as the absence of non-dynamical (`absolute') fields, which would count as a background. The difficulty lies in determining which fields are included in the definition and which not, as the requirement `specified in the KPMs of the theory' can seem arbitrary. Critics could say that too much weight has been put on the semantic presentation of theories, and that we should define more clearly what it means for an object to be `dynamical.'

\subsection{Variational Principles}\label{vp}
Formalising the intuition of what it means to be `dynamical' is what drives the approach for defining background-independence in terms of variational principles, in the recent literature largely originating from \citet{pooley_background_2017}. Pooley's idea is that a field that is dynamical should be subject to Hamilton's principle, i.e.~it should appear as a dependent variable in an action. He argues that what distinguishes GR from background-dependent theories is that its solution space is determined from a generally covariant action by varying it with respect to \textit{all} fields, including the spacetime metric, such that diffeomorphisms are a symmetry group of the action. So more generally, we should find a formulation of the theory in the form of an action principle and check whether this formulation is diffeomorphism invariant. This suggests the following definition for background-independence \citep[p.~28]{pooley_background_2017}:
\begin{definition}[Background-independence in terms of the variation of an action]
\label{def:BIVP}
    A theory is background-independent if and only if its solution space is determined by varying a generally covariant action with respect to all dependent variables that enter the theory. 
\end{definition}
A problem for Definition \ref{def:BIVP} is that also evidently background-dependent theories like special relativity can be derived from varying a generally covariant action, if we add auxiliary fields \citep{sorkin_example_2002}. Thus, if Definition \ref{def:BIVP} is to be of any pratical use, we should restrict to theories that contain only `physical' fields, suggests Pooley. But this notion is again ambiguous.


Another possible pitfall for Definition \ref{def:BIVP} is the question whether it can be applied also to quantum theories, which is important in the context of quantum gravity. So far, we have left the distinction between classical and quantum theories largely implicit. The reason for this was that, even though the definitions of the KPMs and DPMs of the theory need to be modified, the notions of absolute objects and absolute fields straightforwardly carry over to quantum theories \citep[see][§5.1]{read_background_2023}. However, with the definition in terms of variational principles this is not the case, since for quantum theories the solution space is not determined by varying a classical action but must allow for quantum fluctuations. 

Following Read, we define KPMs of a quantum theory to be tuples $\modelthree{M}{F_i}{\hat{O}_i }$, where the $F_i$ represent fixed fields and the $\hat{O}_i$ operator-valued fields. The DPMs of a quantum theory should not be restricted to those satisfying the field equations, since these only give us the tree-level and not the loop contributions. Rather, the relevant equations capturing the dynamics of the theory are the correlation functions for the $\hat{O}_i$. Generally, the correlation functions can be computed using a path integral. In that case, the theory is specified completely by the KPMs as defined above and a path integral, which picks out the DPMs of the theory.\footnote{For quantum theories that do not have a path integral formulation, the DPMs can be picked out directly by the correlation functions, in which case we also need to specify a ground state in the KPMs of the theory \citep{read_background_2023}. However, such theories will not be relevant in this paper.} 

Comparing classical and quantum theories, it is the path integral that takes over the role of the action, since the path integral is used to determine the relevant dynamical equations, i.e.~the correlation functions. The proper analogue of Definition \ref{def:BIVP} for a quantum theory should thus be expressed in terms of the path integral. \citet[p.~101]{read_background_2023} therefore proposes the following modification of Definition \ref{def:BIVP}:
\begin{definition}[Background-independence in terms of the path integral]
\label{def:BIPI}
A quantum theory is background-independent if and only if the solution space is determined by a generally covariant path integral by integrating over all dependent variables, and whose dependent variables represent only physical fields.
\end{definition}
Unfortunately,  Definition \ref{def:BIPI} will suffer from the same difficulties as Definition \ref{def:BIVP}. And perhaps even more: also in cases where a field is considered unphysical on the grounds of its being unobservable in the classical theory, the field may become observable in the quantum theory as fluctuations around the classical solution produce a non-zero effect on the matter and spacetime fields. 

\subsection{Belot's Proposal}\label{belot}
Another approach to defining background-independence is offered by \citet{belot_background-independence_2011}. Belot's starting point is that, generally, whether a theory is background-independent is a matter both of degree and of interpretation. For example, suppose we have a theory where the metric is dynamical but subject to asymptotic boundary conditions: such a theory should perhaps be regarded as being less background-independent than a theory without boundary conditions, but it is certainly more background-independent than a theory in which the metric is not allowed to change at all.\footnote{See also \citet{de_haro_conceptual_2016}} Belot thus argues that  background-independence intuitively comes in degrees, with many theories falling somewhere in between full background-dependence and full background-independence. Also, different formulations of the same theory may be assigned a different degree of background-independence depending on what fields are interpreted as representing the physical content of the theory, and what fields are merely mathematical construction (as we have already seen in Section \ref{vp}). Therefore, background-independence should not be seen as a purely formal notion but as depending on interpretation.

Given these two premises, Belot's idea is that for a theory to be background-independent, physically different DPMs of the theory should describe different spacetime geometries. The question is then what DPMs count as physically inequivalent, and how can we see what geometries they describe. For Belot, evaluating this comes down to identifying what he calls the `physical and geometrical degrees of freedom' of the theory. He defines physical degrees of freedom as those that parametrize the quotient space of DPMs modulo gauge equivalence, where gauge equivalence is defined by the requirement that the theory should be deterministic \citep[see][§5]{belot_background-independence_2011}. Geometrical degrees of freedom are determined by establishing what Belot calls a \textit{geometrization} for a theory: a rule that assigns to every DPM a geometric structure for the manifold of the theory and an equivalence relation for such structures. In most familiar cases, the rule will be to assign a metric for the manifold to every solution and the equivalence relation will be diffeomorphism equivalence. Once a geometrization is given, the geometrical degrees of freedom are identified as those parametrising the set of equivalence classes of geometries. The fact that Belot leaves the details of the geometrization unspecified is precisely the reason why background-independence will be interpretation dependent in his definition. 

With these definitions in hand, Belot proposes the following definition for background-independence:
\begin{definition}[Full background-(in)dependence in Belot's proposal]
\label{def:fullBI}
    A theory is fully background-dependent if it has no geometrical degrees of freedom. A theory is fully background-independent if all physical degrees correspond to geometrical degrees of freedom. 
\end{definition}

As promised, failing the condition for full background-dependence in Definition \ref{def:fullBI} does not imply that the theory is fully background-independent nor vice versa, allowing for different degrees of background-independence in between. To formalize this further, Belot supplements Definition \ref{def:fullBI} with the following definition for near background-(in)dependence:
\begin{definition}[Near background-(in)dependence in Belot's proposal]
\label{def:nearBI}
    A theory is nearly background-dependent if it has only finitely many geometrical degrees of freedom. It is nearly background-independent if it has only finitely many physical degrees of freedom not corresponding to geometrical degrees of freedom.\footnote{Note that this definition in principle allows for a theory to be both nearly background-independent and nearly background-dependent but only if the theory has a finite number of physical degrees of freedom. We will primarily be interested in QFTs, that have an infinite number of physical degrees of freedom.}
\end{definition}
Under this definition, a nearly background-dependent theory could be, for example, a theory in which the metric is fixed to be conformally flat for all solutions. Such a theory would have one geometrical degree of freedom: the conformal factor (but only if we interpret metrics related by a conformal factor as geometrically inequivalent). An example of a nearly background-independent theory is GR with asymptotic flatness imposed: in the natural interpretation, two geometries are considered equivalent when they are related by a diffeomorphism. On the other hand, two solutions should be considered physically equivalent only if they preserve the boundary conditions at infinity. There are thus more physical degrees of freedom than geometrical degrees of freedom. The quotient of the group of diffeomorphisms by the group of diffeomorphisms that act as the identity at spatial infinity gives the Poincaré group, which is ten-dimensional. Hence there are only finitely many physical degrees of freedom that are not geometrical, and the theory is nearly background-independent. Intuitively, Definition \ref{def:nearBI} appears to give the correct verdicts in these examples.

A clear advantage of Belot's definition is that its flexibility in allowing for different degrees widens its range of validity and makes it less prone to fall victim to counter examples. The fact that the degree of background-independence becomes dependent on how we interpret the theory in Belot's definition can be seen as both a strength and a weakness. \citet[p.~33]{read_background_2023} expresses the concern that, if one accepts that there is no a priori ground for saying what fields determine the geometric structure of the theory, one could argue that all fields should be treated on an equal footing, rather than relying on an arbitrary choice of geometrization. On the other hand, Belot's definition does not forbid us to treat all fields on a par---it only asks us to give some interpretation of how the theory translates to a geometric structure, which seems like a rather minimal requirement if a notion of background-independence is to make any sense. That the definition in principle works for any such interpretation makes it more flexible and generally applicable, and perhaps Belot touches upon an important point here that was missing in the other definitions. Making the dependence on the interpretation of the theory explicit makes it more transparent relative to what interpretation a theory is considered background-independent or not. This will prove relevant in the next section, where we will see that the *-algebra is taken to be background-independent precisely because it does not depend on an a priori interpretation in terms of spacetime, but becomes background-dependent if we represent, and then interpret, the algebra as an algebra of Hilbert space operators. Moreover, the semiclassical gravitational setting in which the algebraic formalism of Section \ref{sec:gravdressing} was constructed intuitively counts as somewhat but not fully background-independent, because the theory can describe at most small fluctuations around a fixed, classical solution of the Einstein field equations. So in this context, it also makes sense to speak of degrees of background-independence.

\section{Background-Independence in Algebraic Quantum Gravity}\label{BIAQG}
\label{section_pbi}
According to \citet[p.~6]{witten_background_2025}, the algebraic framework introduced in Section \ref{sectionCLPW} constitutes a background-independent approach to quantum gravity. Our goal in this section is to assess and to qualify this claim. We will relate the algebraic approach to the three notions of background-independence discussed in Section \ref{section_definitions}. However, our aim is not to prove Witten right or wrong with respect to existing definitions of background-independence: as we will argue, these notions are insufficient to fully capture the sense in which the algebraic framework can be more of less background-independent. Rather, we will strive to extract a new notion of background-independence from the algebraic framework, which we argue can be applied to semiclassical or linearized approaches to quantum gravity more generally. 

\subsection{Background-Independence in Semiclassical Gravity}
\label{sec:BIinSCG}
The algebraic framework introduced by CLPW is formulated in a semiclassical gravitational setting. So before evaluating the background-independence of their construction, it may be useful to first note a few more general points regarding semiclassical approaches to quantum gravity. 

Semiclassical gravity means that we treat the gravitational field classically while we take the matter fields to be quantum. This is a good approximation in the limit $G\to 0$, where the metric and the matter fields are decoupled, which allows us to define the quantum matter fields on some fixed solution to the Einstein field equations. Additionally, we can consider small linearized perturbations of the metric and quantize these \citep[see e.g.][]{kiefer_quantum_2007}. Then the quantum fields also include free gravitons. 

Formally, the semiclassical approximation to quantum gravity may be characterized as a perturbative expansion of the Wheeler-DeWitt equation in powers of Newton's constant $G$ \citep[see e.g.][]{singh_notes_1989, ehlers_semiclassical_1994}.\footnote{This is analogous to a Born-Oppenheimer approximation, where $M\sim1/G$ takes the role of the large nucleus mass and the mass of the matter fields $m$ the role of the electron mass.} The leading contribution is of order $1/G$: at this order, one recovers the Hamilton-Jacobi equation for a free gravitational field, which can be used to derive the Einstein equations \citep{gerlach_derivation_1969}. The next order, $\Order{G^0}$, gives the functional Schrödinger equation for quantum fields in curved spacetime. Here the metric obtained at $\Order{1/G}$ can be used as an input to describe the dynamics of quantum fields on a fixed background. Higher-order corrections include backreaction effects of the quantum fields on the metric, such as those described by the semiclassical Einstein field equations: 
\begin{equation}
\label{eq:semiclassicaleinstein}
    G_{\mu\nu} = 8\pi G \expval{T_{\mu\nu}},
\end{equation}
which can be derived from the perturbative expansion of the Wheeler-DeWitt equation at $\Order{G}$. So at this point, the metric becomes dynamical and the DPMs of the theory should be consistent with Eq.~\eqref{eq:semiclassicaleinstein}. If the theory includes free gravitons, these should contribute to the stress-energy tensor, since they are quantum fields.\footnote{To include gravitons, we write the metric that is used in the Wheeler-DeWitt equations as $\metric=\hat g_{\mu\nu} + \sqrt{32\pi G}\,h_{\mu\nu}$, where $\hat g_{\mu\nu}$ is fixed and $h_{\mu\nu}$ satisfies the linearized Einstein equations. $h_{\mu\nu}$ will then appear at the same footing as the matter fields, see \citet{ehlers_semiclassical_1994}. \label{note:gravitons}}

How does this relate to CLPW's algebraic framework? Recall that, as we noted towards the end of Section \ref{section:dS}, in their construction, perturbations of order $\Order{1/G}$ affect the size of the Hilbert space. This reflects the fact that every Hilbert space representation of the algebra is associated to a particular fixed solution of the Einstein field equations. Within every such Hilbert space, states describe $\Order{1}$ perturbations around the input state. So the Hilbert space describes the dynamics of quantum fields on a fixed spacetime background. Backreaction effects are not taken into account in CLPW's construction. 

Before we continue to evaluate the algebraic framework with respect to current notions of background-independence in the philosophical literature, we should emphasize that we should not expect the theory, like any semiclassical theory of gravity, to fully satisfy the conditions for background-independence. These definitions all formulated some way in which the background has to be dynamical. It is clear from the above analysis that a semiclassical theory of gravity can at most describe small fluctuations around some fixed spacetime structure: the background can then be considered dynamical only is a restricted sense. Moreover, CLPW do not explicitly include such dynamical perturbations of the spacetime in their framework, so this cannot be the reason why Witten claims that the construction is background-independent. Rather, there seems to be some different notion of background-independence at play here: the fact that the construction of the algebra does not depend on a specific spacetime formulation and can thus in principle be used for any background, in the sense of (ii) formulated in Section \ref{sec:introduction}.  This suggests a different notion of background-independence for semiclassical quantum gravity more generally: the fact that we can perturb around any classical solution to the Einstein equation.

\subsection{Relating to Existing Notions of Background-Independence}
\label{sec:existingdef}
To relate CLPW's algebraic framework to the three notions of background-independence discussed in Section \ref{section_definitions}, we have to translate the framework into the semantic conception of theories used there, in terms of KPMs and DPMs. Since every state on the algebra is associated to a specific spacetime background, observer worldline, and expectation value for the observables of the theory, we can interpret the states (in combination with the algebra) as being the models $\modelthree{M}{\gamma}{\phi}$ of the theory. However, identifying the KPMs and DPMs that define the theory is not straightforward. Recall that every state defines a Hilbert space representation for the algebra via the GNS-construction, and that states associated to unitary equivalent Hilbert spaces can be interpreted as corresponding to equivalent choices of $M$ and $\gamma$. For simplictly, let us assume that in every such Hilbert space we can define a Hamiltonian and a path integral. Then there seem to be two options here:
\begin{enumerate}[label=(\Alph*)]
    \item The KPMs of the theory are the states living in a single Hilbert space for the theory with an associated path integral. The DPMs are the states within that Hilbert space for which the correlation functions agree with the path integral.\footnote{Or more generally, since not all QFTs have a path integral formulation, the DPMs can be determined directly by a set of correlation functions, in which case we also need to specify a state. If there is a path integral formulation, these to approaches are equivalent. See also \citet[§5.1.2 and Footnote 8]{read_background_2023}.\label{note:pi}} 
    \item The KPMs of the theory are all states belonging to the different Hilbert spaces on which the algebra can be represented. DPMs are the states for which the correlation functions agree with the path integral in those Hilbert spaces. 
\end{enumerate}
This distinction is important, as options (A) and (B) will lead to different evaluations of the theory's background-independence. Which option appears more natural? Strictly speaking, dynamics is defined only within a particular Hilbert space, with respect to a specific choice of $M$ and $\gamma$. So from this perspective, (A) seems more natural. On the other hand, if we take the underlying *-algebra (or operator product algebra, see Footnote \ref{note:opa}) to be the defining structure for the theory, rather than one of its particular Hilbert space representation, then the possible models of the theory are not limited to a single Hilbert space. From that perspective, (B) seems more natural. In other words, the question is what we define to be `the theory'---and as we will see, this issue is not only logical, but has physical consequences.

When choosing option (A), the theory appears to be background-dependent, since the theory is limited to a single Hilbert space associated to a particular choice of background. Indeed, when considering Definition \ref{def:BIAF} in terms of absolute fields, the theory comes out as background-dependent: all KPMs of the theory involve the same metric field, so this metric field will match the definition of an absolute field. Also in Definition \ref{def:BIPI}, the quantum version of background-independence based on variational principles, the theory is background-dependent: because the metric is fixed, it cannot be integrated over in a generally covariant path integral to obtain the Hilbert space. Finally, also Belot's definition (Definition \ref{def:fullBI}) judges the theory to be fully background-dependent: we assign to every state the same geometric structure, hence there are no geometric degrees of freedom that can correspond to physical degrees of freedom. For option (A), the theory is thus unambiguously background-dependent according to existing definitions. Perhaps this is not surprising, since it describes a QFT on a {\it fixed} curved background.

As an aside, we may note that things become more interesting when we include $\Order{G}$ backreaction effects. In that case, the theory describes small fluctuations around a fixed background, of the form:
\begin{equation}
\label{eq:metricperturbations}
    \metric = \hat{g}_{\mu\nu} + \delta g_{\mu\nu},
\end{equation}
where $\hat{g}_{\mu\nu}$ is a fixed solution and $\delta\metric$ is now dynamically coupled to the matter fields through Eq.~\eqref{eq:semiclassicaleinstein}. (Additionally,  Eq.\eqref{eq:metricperturbations} may also include $\Order{\sqrt{G}}$ graviton contributions, see Footnote \ref{note:gravitons}.) This theory will still be background-dependent according to absolute object-type definitions, since the $\hat{g}_{\mu\nu}$ is both an absolute object and an absolute field. However, according to Definition \ref{def:BIPI} based on variational principles, the verdict is different. Recall that in Section \ref{section:dS}, we discussed Witten's conjecture that for any spacetime background a Hartle-Hawking state can be constructed, suitably generalized to include an observer, by an Euclidean path integral which includes the metric as a dynamical field and takes the background metric $\hat{g}_{\mu\nu}$ as a boundary condition. The Hartle-Hawking state then determines the Hilbert space via the GNS construction. Supposing this conjecture is true, the theory thus comes out as background-independent in Definition \ref{def:BIPI}: its solution space is completely determined by a generally covariant path integral, where we integrate over all dependent variables. The background $\hat{g}_{\mu\nu}$ does not need to be integrated over, since it only appears as a boundary condition. Something similar happens in Belot's definition. By virtue of Eq.~(\ref{eq:semiclassicaleinstein}) and (\ref{eq:metricperturbations}), all physical degrees of freedom correspond to geometrical degrees of freedom, and the theory is thus fully background-independent according to Definition \ref{def:fullBI}. This is perhaps surprising: since the dynamics of the geometry is restricted to small perturbations, we would expect the theory to be nearly background-dependent. But the truth is that Belot's definition does not care whether the fluctuations are small or not, it only cares whether distinct physical configurations lead to distinct geometries, and this is not affected by taking $G$ to be small. 

Next, we consider what happens if we choose option (B). Now, the metric field varies across the KPMs and DPMs of the theory. Therefore, it is no longer an absolute object nor an absolute field. The theory thus comes out as background-independent according to Definition \ref{def:BIAF}. However, in Definition \ref{def:BIPI} based on variational principles the theory remains background-dependent: the path integral corresponds to a particular Hilbert space representation of the algebra.\footnote{Or more generally, a set of correlation functions for the fields associated to a particular state (see also Footnote \ref{note:pi} above).} Therefore, it is not clear how a single path integral could determine a space of DPMs which consists of states belonging to an infinite number of inequivalent Hilbert spaces. Instead, different (irreducible) representations of the algebra can be interpreted as superselection sectors of the theory.\footnote{This idea originates from \citet{doplicher_fields_1969}; see also \citet{halvorson_algebraic_2007} for a philosophical discussion.} Superselection sectors are not summed over in the path integral, but can be interpreted as self-contained theories. The fact that absolute object-type definitions and those based on variational principles give different verdicts reflects the different principles underlying these definitions: in absolute object-type definitions, the spacetime structure is required to vary across the DPMs of the theory, but it does not matter how or why. By contrast, in definitions based on variational principles, the background is only considered `dynamical' if it is integrated over as a dependent variable in the path integral. In option (B), the spacetime structure is allowed to vary across the DPMs of the theory, but not determined `dynamically' in this sense. Rather, it is determined by choice. 

The different verdicts suggest that there might be an intermediate form of background-independence. Indeed, this is what we find in Belot's definition. Even though there is no straightforward `geometrization' for the theory, it is clear that there will be more physical degrees of freedom than geometrical degrees of freedom. Namely, within every Hilbert space there are different physical degrees freedom corresponding to a single background geometry. Therefore, we may say that the theory is somewhere in between full background-independence and full background-dependence (but note that Definition \ref{def:nearBI} for near background-independence and near background-dependence is not applicable either, because it assumes finite numbers of degrees of freedom, while the number of physical degrees of freedom not corresponding to geometrical degrees of freedom is here infinite, and there is also an infinite number of geometrical degrees of freedom). 

We thus see that options (A) and (B) lead to different assessments of the theory's background-independence, except for the two definitions based on variational principles, according to both of which the theory comes out as background-dependent. That the theory is not background-independent in case (A) is not surprising and in fact, this is also emphasized by \citet{witten_background_2025}.  He argues that to get a background-independent theory, we should opt for (B) and view the algebra of observables as an abstract *-algebra rather than as an algebra of operators on a Hilbert space. Then the algebra defines an underlying structure of the theory, which does not rely on the commitment to a specific spacetime structure. Of course, to be able to do any interesting calculation, we still have to pick a background state and close the algebra to a von Neumann algebra in the associated GNS Hilbert space. But the point is to see the *-algebra, rather than the Hilbert space, as the underlying structure that defines the theory.  Such an approach may be motivated by the observation that, even for a fixed background, there may in principle be an infinite number of inequivalent Hilbert space representations of the algebra \cite{summers_stone-von_2001,ruetsche_interpreting_2002}. 

Still, existing definitions of background-independence do not give an entirely satisfactory verdict in case (B). Definition \ref{def:BIAF} judges the theory to be completely background-independent if we do not restrict our attention to a particular representation of the algebra, but we thereby fail to take into account that the background-independent *-algebra is not itself a complete theory, and that any actual calculation will depend on an arbitrary choice of background rather than determine this background dynamically. Intuitively, the semiclassical algebraic framework should therefore be assigned a weaker form of background-independence than, say, GR. Besides, the absolute object-type definition judges the framework to be background-independent on account of there not being a field which is invariant under the dynamics, while the *-algebra intuitively counts as background-independent on account of its structure being the same for different choices of background, which is conceptually different.

Belot's definition does give the intuitively right verdict that the theory is somewhere in between full background-independence and full background-dependence. However, the language in terms of geometric and physical degrees of freedom that Belot's approach uses seems ill-suited to the algebraic framework. Belot's definition depends on a geometrization of the theory, but the point of (B) is precisely that we do not commit to a specific spacetime interpretation of the *-algebra but leave all options open. As a result, it is difficult to quantify geometrical and physical degrees of freedom, and the sense in which the framework is somewhat but not fully background-independent cannot be made precise. 

\subsection{An Algebraic View on Background-Independence}
\label{sec:newnotion}
The upshot of the above discussion is that the algebraic framework of CLPW appears to be background-independent in a sense that is not well captured by existing notions of background-independence in the philosophical literature. These notions all defined some sense in which the background was supposed to be dynamical for the theory to be background-independent. This does not seem the kind of background-independence that \citet[p.~6]{witten_background_2025} refers to when he writes that the algebraic construction formulated in Section \ref{sec:gravdressing} ``hopefully sounds `background-independent', since we described it without picking a background." Rather, Witten seems to suggest that the algebraic construction is background-independent because it does not rely on any spacetime structure at all. 

So what does Witten mean? Crucially, he emphasizes that the algebra is not background-independent if we interpret it as an algebra of bounded operators on a Hilbert space: as discussed in Section \ref{section:whyaqft}, the construction of the Hilbert space depends on a specific choice of background. However, if we interpret the algebra as an abstract *-algebra, independent of a Hilbert representation, the algebra is also not dependent on any spacetime background. The operators in the algebra are then associated only to some proper time interval along the trajectory of the observer, and interpreted as the observables accessible to the observer in that interval. As such, the *-algebra defines an underlying structure for the theory that is the same for --- and in that sense independent of --- any choice of background, encoding its causal structure through a set of operator relations. The operators can be given a spacetime interpretation by embedding the observer's worldline into a spacetime background, which is equivalent to considering a specific state or Hilbert state representation of the algebra. 

There are two points to be made here. First, the conceptual difference between the usual way in which background-independence is understood in the philosophical literature and the notion of background-independence suggested by Witten is that in the latter the spacetime background is not required to be dynamical. Rather, in the *-algebra there is no reference to a spacetime structure at all. It is true that Witten interprets different states on the algebras to be associated to different spacetime backgrounds, and that he thus in that sense presupposes a `dynamical' background. But the crucial difference is that the background is determined arbitrarily and not through a dynamical coupling to physical degrees of freedom, as in the Einstein equations in GR. On the contrary, dynamics is defined here only for a fixed choice of background. Background-independence then means that  (a) we do not rely on a specific background, 
and (b) we  
are free to choose one. 

Second,  and relatedly,  the notion of background-independence that Witten suggests is only a partial background-independence. The *-algebra should not, on itself, be interpreted as a complete theory. Even though we can assign an expectation value to operators in a particular state, there is no dynamics that can be used to make predictions. Besides, it is not always clear whether the *-algebra includes the full set of the physically interesting observables \citep[see][§4.5]{wald_quantum_1994}. Instead, we should see the *-algebra as a basic structure on which we can construct different theories. While the notion of background-independence that Witten suggests only applies to this basic structure, and not to the complete theory,  the notion includes that some full set of observables exists: and this explains the role of the von Neumann algebras. 

These points suggest the following definition:
\begin{definition}[Background-Independent Algebra]
\label{def:bialgebra}
    An algebra is background-independent iff (a) this algebra can be interpreted as an abstract operator algebra that does not rely on a specification of the structure of spacetime,  and (b) it admits the specification of such a structure. 
\end{definition}
So this is a new notion of background-independence specifically for algebras (as against theories), along the lines of point (ii) in Section \ref{sec:introduction}. In particular, as we have seen, this means that the algebra should not be interpreted as the algebra of operators on a particular Hilbert space. The algebra has to be representation-independent to be background-independent, yet it still must admit the specification of a Hilbert space (for example via the GNS construction). In this sense, we can say that the algebra admits a spacetime interpretation, even though it does not depend on a specific spacetime.\footnote{But representation independence is not a sufficient condition for background-independence of the algebra, since there can be many unitarily inequivalent Hilbert space representations of the algebra even for a fixed background.} As we argued in Section \ref{belot}, the flexibility of the definition with respect to different interpretations of the spacetime structure of the theory both makes this necessary dependence on the interpretation more transparent and ensures that the definition is more generally applicable, which supports the idea that background-independence comes in degrees, similarly as in Belot's proposal.

A natural question to ask is what the background-independence of the algebra means for the background-independence of the full spacetime theory. The advantage of the background-independent algebra is that it provides a standard building block to construct theories in different backgrounds, and this is relevant for semiclassical approaches to quantum gravity. As we noted above, in semiclassical gravity we do not expect the background to be truly dynamical, except for small perturbations around a fixed solution. So what form of background-independence can we expect in this case? Leaving such fluctuations aside for now, a semiclassical theory that makes no a priori assumption about the background structure, except that it must exist, is intuitively more background-independent than a semiclassical theory that can be formulated only on, say, spacetimes with a negative cosmological constant. In other words, the most `background-independent' that we can hope for in a semiclassical theory is that we can formulate the theory for \textit{any background}. Here, by `background' we mean a fixed solution to the classical Einstein equations. The background-independent algebra provides a basis to realize this: we associate states on the algebra to different spacetime backgrounds. Every such state defines a Hilbert space representation for the algebra via the GNS-construction, and we can close the algebra in this representation to a von Neumann algebra, define dynamics, etc. The result is a QFT on a fixed classical solution to the Einstein equation. Since the underlying algebraic structure does not rely on a spacetime formulation at all, it can in principle  be used to formulate a theory on any background. The background-independence of the algebra thus suggests a notion of background-independence of semiclassical gravitational theories, that can be summarized by the following definition:
\begin{definition}[Semiclassical Background-Independence]
\label{def:SCBI}
    A quantum theory is semiclassicaly background-independent iff in the limit $G\to 0$ it can be formulated for any solution to the classical Einstein equation.
\end{definition}
A background-independent algebra is not only useful for the construction of semiclassically background-independent theories, but can also be used to derive general properties of the theory that hold independently of the choice of background. For example, in CLPW's construction discussed in Section \ref{sec:gravdressing}, we learned that the dynamics of the field operators along the worldline of the observer takes the form of the familiar Schrödinger equation through a Page-Wootters-like mechanism (see Eq.~(\ref{eq:schrödingerequation})). So, even though the Hamiltonian changes, the dynamics that emerges from the perspective of the observer is captured by the same equation. Also, as we discussed in Section \ref{section:gen}, the hypothesis is that for any background the algebra is of Type II and includes a notion of entropy and density matrices.

So far we have left backreaction effects completely out of the picture and worked in the strict $G\to0$ limit. However, we may now wonder how robust the background-independent algebra and the general statements that we derived from it are against perturbative corrections in $G$. When $G$ is non-zero, both the observer and the quantum fields backreact on the metric, and are thus coupled to each other. This means that the Hamiltonian constraint in Eq.~(\ref{eq:htc}) that we used to gravitationally dress the algebra to the worldline of the observer is no longer a simple sum. So the algebra is potentially deformed by $\Order{G}$ corrections. However, as \citet{witten_algebras_2024} argues, the important point here is that the algebra that we constructed in the $G\to0$ limit is a \textit{factor}, meaning that it has a trivial centre---which has a crucial consequence: factors are rigid, i.e.~they have no infinitesimal deformations (in this sense, a factor can be seen as the von Neumann analogue of a simple Lie group). Therefore, perturbative corrections in $G$ can modify the algebra up to isomorphism. That is, they can change the commutation relations of the operators, but not the intrinsic structure of the algebra as captured by its Type. The conclusions that we have drawn are thus expected to hold in the perturbative regime. However, as we noted in Section \ref{sec:gravdressing}, we have to be careful in spacetimes with a positive cosmological constant, where small perturbations can grow indefinitely.

What does the inclusion of $\Order{G}$ corrections mean for the background-independence of the (semiclassically background-independent) theory? By turning on $G$, the background actually becomes dynamically coupled to the quantum fields through Eq.~(\ref{eq:semiclassicaleinstein}), albeit only in the form of small fluctuations around a fixed solution. Thus the theory now seems to combine two notions of background-independence: (1) that of Definition \ref{def:SCBI}, i.e.~that it can be formulated for any background, and (2) that of a dynamical background. Sense (1) remains relevant because the dynamics of (2) remain close to a particular solution of the Einstein equation: the theory should be considered less or more background-independent depending on whether it describes small perturbations around a only single solution or around any solution. Besides, as we discussed in Section \ref{sec:BIinSCG}, the fact that the dynamics is limited to small perturbations of a fixed background means that the theory should only be considered somewhat but not fully background-independent in the sense of (2). Belot's approach to background-independence is the only approach that we saw that allows for intermediate degrees. However, as we saw in Section \ref{sec:existingdef}, Belot's definition is not well-suited for perturbative quantum gravity, as the degree of background-independence is related only to the number of physical and geometrical degrees of freedom and not the relative size of their fluctuations. Therefore, to better capture this notion of \textit{perturbative background-independence} inspired by the algebraic framework, we propose the following definition.
\begin{definition}[Perturbative background-independence]\label{def:pbi}
    A theory is perturbatively background-independent iff:
    \begin{enumerate}[label=(\arabic*)]
        \item The theory is semiclassically background-independent in the limit $G\to0$. 
        \item The theory describes perturbations around a fixed classical solution to the Einstein equations for small but finite values of $G$.
    \end{enumerate}
\end{definition}
A theory that is perturbatively background-independent is thus more background-independent than a theory that is only semiclassicaly background-independent, but less background-independent than a theory in which the background is fully dynamical, as captured by Definitions \ref{def:BIAF}, \ref{def:BIPI}, and \ref{def:fullBI}. CLPW’s algebraic framework, as we formulated it in Section \ref{sec:gravdressing} is semiclassically background-independent but \textit{not} perturbatively background-independent, because it was constructed only in the strict semiclassical limit $G \to 0$. However, as we argued above, we expect that the construction is robust against perturbative effects in $G$. If we include such effects, the theory becomes perturbatively background-independent.

Definition \ref{def:pbi} of perturbative background-independence can be compared to the minimalist conception of background-independence defined by \citet{de_haro_conceptual_2016}. In this minimalist conception, which was formulated in the context of the AdS-CFT duality, a theory is background-independent if the background is determined dynamically up to boundary conditions, which contrasts with the ‘extended sense’ in which also the boundary conditions have to be determined dynamically. The similarity with the notion of perturbative background-independence as defined in Definition \ref{def:pbi} is that, in both cases, we first have to make an arbitrary choice of background representation or boundary conditions, and then the theory describes the dynamics of the background relative to this choice. A difference is that in Definition \ref{def:pbi}, the dynamics of the background are limited to small perturbations of $\Order{G}$, while in the minimalist conception fluctuations of the background can be of arbitrary size but have to reduce to the boundary condition at asymptotic infinity.

\subsection{Why Should We Care About Algebraic Background-Independence?}
In the last two sections, we saw that conventional notions of background-independence are not well-suited for the algebraic approach and instead formulated a novel notion that seems to best capture what Witten means in this context. Now, let us take a step back and see how this novel notion connects to the original motivations for background-independence, and why we should still care for it under that name. 


As noted in Section \ref{sec:introduction}, a common thread in different approaches to quantum gravity is that they seek to extract what is `essential' in GR, and it is this intuition that definitions of background-independence aim to formalise: that spacetime itself is a dynamical field, implying that there can be no prior dependence on a fixed geometry. However, practically, we often resort to semiclassical approximations in which the dynamics of the background are frozen. In this context we certainly cannot expect full background-independence in the dynamical sense. The question then becomes: is there some residual form of background-independence at the semiclassical level that avoids an arbitrary dependence on a pre-selected background, without requiring the background to be fully dynamical? 

Such a form of background-independence arises naturally in the algebraic construction of Section \ref{sectionCLPW} at the level of the *-algebra: an underlying algebraic framework that is independent of a particular spacetime realisation (cf. Bergmann's idea of a `skeletonized' theory). This notion captured in Definition \ref{def:bialgebra} can be used to to study what is common to semiclassical theories formulated in different spacetimes. The notion of algebraic background-independence thus partially preserves the original motivation for background-independence from GR --- avoiding arbitrary dependence on a pre-selected background --- within the limitations of the semiclassical approximation. Moreover, point (b) in Definition \ref{def:bialgebra} secures that any background-independent algebra can be used to constructed a spacetime theory. Although the underlying framework does not depend on a background, it can be assigned a class of spacetimes that are consistent with it. Also in this sense, the algebraic notion of background-independence is not so remote from the usual one; and, since it is prior to the specification of the geometry, we argue that it is more fundamental.

Definitions \ref{def:SCBI} and \ref{def:pbi} extend the same intuition to the level of theories. Together, they provide conceptual tools to analyse the scope and limitations of semiclassical theories. And it is meaningful to do so under the name of background-independence, as this allows us to situate such theories somewhere between full background-dependence and full background-independence in the dynamical sense of GR --- similar in spirit to Definition \ref{def:nearBI} in Belot's proposal, which however, as was argued in Section \ref{sec:newnotion}, is not well-suited for the semiclassical context. The alternative, concluding that background-independence is simply inapplicable to semiclassical theories, would leave us without the conceptual tools to distinguish, for instance, a theory formulated only on de Sitter space from one that is equally valid on any background --- a distinction that is both technically meaningful and physically important.

The potential use of an algebraic notion of background-independence can be illustrated by relating it back to contemporary developments in quantum gravity. At the moment, AdS-CFT provides one of the most powerful frameworks to study quantum gravity in a concrete setting. AdS-CFT is not (fully) semiclassically background-independent in the sense of Definition \ref{def:SCBI}, as it is limited to solutions of the Einstein equations with a negative cosmological constant. It can be difficult to assess whether the things that we learn in AdS-CFT teach us about gravity more generally, or are specific to the AdS setting. A possible strategy to generalize AdS-CFT to arbitrary values of the cosmological constant is to look for an underlying structure that does not depend on the specifics of the background. As we have seen, such a structure is naturally formulated in an algebraic language. In a recent paper, \citet{liu_towards_2025}, argues that the algebraic description of AdS-CFT can be generalized by considering different states on the same algebra that correspond to different geometries, including de Sitter-type spacetimes with a positive cosmological constant, by showing the viability of `baby universes,' and in particular the Antonini-Sasieta-Swingle and Maldacena-Maoz geometries. \citet{kudler-flam_emergent_2025} show that in the semiclassical limit the Antonini-Sasieta-Swingle geometry corresponds to a mixed state in the CFT. These results are promising for the establishment of a connection between different developments in AdS-CFT and de Sitter-type spacetimes, like that of CLPW, into a single algebraic framework. As \citet{liu_towards_2025} puts it, such an algebraic framework would offer ``a unified perspective on quantum gravity, independent of asymptotic structures or the sign of the cosmological constant." The strength of this framework is then precisely that it is background-independent, not in the dynamical sense, but in the algebraic sense defined in this paper, providing a unified framework that works for any sort of background.

\section{Conclusion}
\label{sec:conclusion}
In this paper, we have discussed a novel approach to constructing operator algebras in quantum gravity and have assessed its background-independence. The approach consists of a procedure that `gravitationally dresses' the operators to the worldline of an observer. The upshot is that this algebra can be interpreted as the algebra of observables that the observer has access to along their worldline. If the algebra is interpreted as an abstract operator product algebra or a *-algebra, rather than as an algebra of Hilbert space operators, it is independent of the spacetime background on which the observer travels, and is thus said to be `background-independent.' States on the algebra can then be interpreted as corresponding to different spacetime backgrounds. 

We compared the notion of background-independence that arises in the algebraic approach to three main existing notions in the philosophical literature---absolute object-type definitions, definitions based on variational principles, and Belot's proposal---and found that it is different from them. These existing notions all require that the theory is background-independent if the spacetime background is dynamical in some way. By contrast, the algebraic approach works in a semiclassical gravitational setting where the dynamics of the background are frozen. So the algebra is not background-independent in the sense of a dynamical background. Rather, the presumed background-independence of the *-algebra refers to its {\it independence from a particular Hilbert space representation or state}, which corresponds to a particular background for the algebra. In other words, the *-algebra is taken to be background-independent because it does not rely on a spacetime formulation at all. We used this to define a more primitive notion of background-independence within the semiclassical setting, that preserves part of its original motivation: avoiding arbitrary dependence on a pre-selected background. The relevance of this notion is that, if the algebra is background-independent in this sense, it means that the algebra can be used to derive general properties of semiclassical gravity in any background. These properties are expected to be robust against perturbative corrections in $G$. As such, the background-independent algebra provides a sort of `skeletonized' structure for the construction of semiclassical gravitational theories in different backgrounds, similar in spirit to early approaches to quantum gravity.

Building on this notion of algebraic background-independence, we defined a theory to be semiclassically background-independent if in the limit $G\to0$ it can be formulated for any solution of the Einstein field equations, and perturbatively background-independent if it also describes perturbations around such a solution for small but finite values of $G$. These definitions are relevant for the qualification of background-independence of theories in the semiclassical regime. If a theory is semiclassically background-independent according to our definition, it means that it provides a unified framework for studying semiclassical gravity in any background, and thus brings us closer to a full theory of quantum than a theory that is restricted to a certain class of geometries. Our definition of perturbative background-independence tells us whether the theory is also expected to be robust against pertubative effects in $G$. In other words, our definitions for semiclassical and perturbative background-independence are useful for assessing the range of validity of semiclassical gravitational theories.

\subsubsection*{Acknowledgements}
The authors thank Enrico Cinti, Niels Martens, James Read, Antony Speranza, and Erik Verlinde for discussions and comments on the text, and are grateful to the E.W. Beth Foundation grant KNAW WF/2531-08 for financial support. LV also thanks the audiences in Oxford and Amsterdam where this work has been presented. LV's work is funded by the NWA ORC programme `Emergence at All Scales' of the Dutch Research Council (NWO). SDH's work is supported by the John Templeton Foundation grant 63670 on `Emergence of time from chaos in operator algebras'. SDH also thanks Jesus College, Oxford for their hospitality and support during the completion of this paper.

\bibliographystyle{apalike}
\bibliography{bibliography}

\end{document}